\documentclass[aps,prl,twocolumn,showpacs,superscriptaddress]{revtex4-2}
\usepackage{dcolumn}% Align table columns on decimal point
\usepackage{graphicx}% Include figure files
\usepackage{array}
\usepackage[utf8x]{inputenc}
\usepackage[dvipsnames]{xcolor}
\usepackage{float}
\usepackage{natbib}
\usepackage{bm}% bold math
\usepackage[T1]{fontenc}
\usepackage{mathptmx}
\usepackage{physics}
\usepackage{amssymb}
\usepackage{amsmath}
\usepackage{nicefrac}
\usepackage{todonotes}
\usepackage{array}
\usepackage{makecell}
\usepackage{csquotes}
\usepackage{upgreek}
\usepackage{mathtools, nccmath}
\usepackage{lipsum}
\usepackage[normalem]{ulem}
\usepackage{dcolumn}% Align table columns on decimal point
\usepackage{xspace}
\usepackage{chemformula} %chemical formulas
\newcommand{\VNS}{\ch{V_{1/3}NbS2}\xspace} 
\usepackage{times} %times new roman
\usepackage{tabularx}
\usepackage{mathtools} 
\usepackage{xcolor}
\usepackage{amsmath} 
\usepackage[version=4]{mhchem}

\usepackage{chngcntr}
\usepackage{hyperref}% add hypertext capabilities

\begin{document}

% ---------- MAIN ----------
\preprint{APS/123-QED}

\title{Stacking-dependent anisotropic altermagnetism in V$_{1/3}$NbS$_2$} 
%\date{December 2021}

\author{Chris J. Lygouras}
\email{chrislygouras@gmail.com}
\affiliation{Institute for Quantum Matter and William H. Miller III Department of Physics and Astronomy, Johns Hopkins University, Baltimore, Maryland 21218, USA}

\author{Nathan Prouse}
\affiliation{Institute for Quantum Matter and William H. Miller III Department of Physics and Astronomy, Johns Hopkins University, Baltimore, Maryland 21218, USA}

\author{Jack H. Drouin}
\affiliation{Institute for Quantum Matter and William H. Miller III Department of Physics and Astronomy, Johns Hopkins University, Baltimore, Maryland 21218, USA}

\author{Youzhe Chen}
\affiliation{Institute for Quantum Matter and William H. Miller III Department of Physics and Astronomy, Johns Hopkins University, Baltimore, Maryland 21218, USA}
\affiliation{Department of Physics, University of California, Berkeley, CA, 94720, USA}
\affiliation{Material Sciences Division, Lawrence Berkeley National Lab, Berkeley, CA, 94720, USA}

\author{Laura Garcia-Gassull}
\affiliation{Institut f{\"u}r Theoretische Physik, Goethe-Universit{\"a}t Frankfurt, Max-von-Laue-Strasse 1, 60438 Frankfurt am Main, Germany}

\author{Zili Feng}
\affiliation{Department of Physics, University of Tokyo, Bunkyo-ku, Tokyo 113-0033, Japan}
\affiliation{Institute for Solid State Physics, University of Tokyo, Kashiwa, Chiba 277-8581, Japan}

\author{Mingxuan Fu}
\affiliation{Department of Physics, University of Tokyo, Bunkyo-ku, Tokyo 113-0033, Japan}
\affiliation{Institute for Solid State Physics, University of Tokyo, Kashiwa, Chiba 277-8581, Japan}

\author{L{\"u} Fang}
\affiliation{Department of Physics, University of Tokyo, Bunkyo-ku, Tokyo 113-0033, Japan}

\author{Alexander I. Kolesnikov} 
\affiliation{Neutron Scattering Division, Oak Ridge National Laboratory, Oak Ridge, TN 37831, USA}

\author{Christina Hoffman}
\affiliation{Neutron Scattering Division, Oak Ridge National Laboratory, Oak Ridge, TN 37831, USA}

\author{Yiqing Hao}
\affiliation{Neutron Scattering Division, Oak Ridge National Laboratory, Oak Ridge, TN 37831, USA}

\author{Huibo Cao}
\affiliation{Neutron Scattering Division, Oak Ridge National Laboratory, Oak Ridge, TN 37831, USA}

\author{Maxime A. Siegler}
\affiliation{Department of Chemistry, Johns Hopkins University, Baltimore, Maryland, 21218, USA.}

\author{Robert J. Birgeneau}
\affiliation{Department of Physics, University of California, Berkeley, CA, 94720, USA}
\affiliation{Material Sciences Division, Lawrence Berkeley National Lab, Berkeley, CA, 94720, USA}

\author{Roser Valent{\'i}} 
\affiliation{Institut f{\"u}r Theoretische Physik, Goethe-Universit{\"a}t Frankfurt, Max-von-Laue-Strasse 1, 60438 Frankfurt am Main, Germany}
\affiliation{Canadian Institute for Advanced Research (CIFAR), Toronto, Ontario M5G 1M1, Canada}

\author{Satoru Nakatsuji}
\affiliation{Institute for Quantum Matter and William H. Miller III Department of Physics and Astronomy, Johns Hopkins University, Baltimore, Maryland 21218, USA}
\affiliation{Department of Physics, University of Tokyo, Bunkyo-ku, Tokyo 113-0033, Japan}
\affiliation{Institute for Solid State Physics, University of Tokyo, Kashiwa, Chiba 277-8581, Japan}
\affiliation{Trans-scale Quantum Science Institute, University of Tokyo, Bunkyo-ku, Tokyo 113-0033, Japan}
\affiliation{Canadian Institute for Advanced Research (CIFAR), Toronto, Ontario M5G 1M1, Canada}

\author{Collin L. Broholm}
\affiliation{Institute for Quantum Matter and William H. Miller III Department of Physics and Astronomy, Johns Hopkins University, Baltimore, Maryland 21218, USA}
\affiliation{NIST Center for Neutron Research, Gaithersburg, Maryland 20899, USA}
\affiliation{Department of Materials Science and Engineering, Johns Hopkins University, Baltimore, Maryland 21218, USA}

\begin{abstract} 
    We report profound impacts of the stacking sequence of triangular lattices of magnetic transition metal ions intercalated between the layers of the van der Waals material NbS$_2$. Using single crystal x-ray and neutron diffraction, and transport and magnetization measurements, we show there are two distinct polytypes of $\rm V_{1/3}NbS_2$ with disparate easy axes of magnetization and different anomalous Hall responses. Self-consistent analysis of inelastic neutron scattering data provides evidence for oscillatory RKKY interactions that extend to 1 nm and stabilize quasi-collinear A-type altermagnetic orders in both polytypes though with perpendicular easy axes. The detailed stacking sequence of a bulk polytype crystal dramatically impact its macroscopic anomalous Hall response and magnetism, which suggests a new path to engineer the bulk properties of a layered three dimensional solid.
\end{abstract} 
\date{\today}
\maketitle 
%\onecolumngrid 

%\section{Introduction} 
Van der Waals (vdW) solids comprise a class of materials where electrically neutral 2D crystalline layers are held together by fluctuating dipolar interactions. They host an array of exceptional physical phenomena, including charge density waves with strong electronic correlations \cite{Moncton1977, Chen2016, Hwang2024}, superconductivity \cite{Morosan2006, Guguchia2017, Heil2017, Witteveen2021}, and magnetism \cite{Guguchia2018, Burch2018, Pasco2019, Ray2025, Razpopov2025}. The weak inter-layer coupling allows them to be mechanically exfoliated toward the two-dimensional limit, where they can host exotic electronic phenomena that cannot be realized in the bulk. For example, the magnetic order can be tuned by the crystal structure and stacking arrangement \cite{Sivadas2018, Pasco2019, Pinto2021, Yang2023}, or by applied electric fields \cite{deng_gate_tunable_2018, Xu2018, Zhang2025}. In addition, net ferromagnetism (FM) can be generated in odd-layer antiferromagnets (AFM) due to uncompensated moments \cite{Fei2018, Yang2021}. This has lead to enhanced Berry curvature manifesting in anomalous or quantized conductivity \cite{Deng2020, Gao2023, Fox2024}. These examples highlight the versatility and tunability of correlated phases in the two-dimensional limit. 

\begin{figure}
    \centering
    \includegraphics[width=\linewidth]{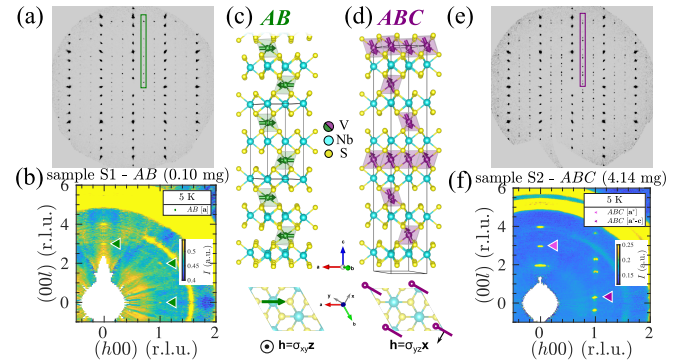}
    \caption{ \textbf{Structures for two \VNS polytypes.} (a) Reconstructed reciprocal-lattice $(h0l)$ slice from SCXRD (at 213~K) and (b) from neutron diffraction (at 5~K) in the $(h0l)$ scattering plane for the (c) $AB$ polytype. For the (d) $ABC$ polytype, we again show the $(h0l)$ slice (e) from SCXRD and (f) from neutron diffraction. Throughout, the unit cell is shown in black borders; the colored arrows correspond to the vanadium spins; and the black arrows indicate the Hall vectors associated with minute spin canting and anomalous Hall effect.} 
    \label{fig:VNS_structure} 
\end{figure} 

Among bulk vdW materials, transition-metal dichalocogenides (TMDs) are of interest due to their structural similarity to graphene, with the honeycomb layers weakly coupled by vdW interactions into a variety of three-dimensional stacking arrangements \cite{Beal1979, Kolobov2016}. Magnetism can be introduced by incorporating magnetic transition metals in the vdW gap \cite{Parkin1983, Park2023, Ray2025, Zhou2025}. The intercalated \ch{$T$_{1/3}NbS2} with $T$ a transition metal ion, features distinct $A$ and $B$ site triangular lattices on either side of the host $2H$-stacked \ch{NbS2} layers, resulting in a $\sqrt{3}\times \sqrt{3}$ superlattice that breaks inversion symmetry (Fig.~\ref{fig:VNS_structure}). Charge transfer between the intercalant and NbS$_2$ strengthens interplanar interactions and results in 3D magnetism. 
%Because magnetic ions are believed to interact via coupling with conduction electrons in a Ruderman-Kittel-Kasuya-Yosida (RKKY) exchange process \cite{edwards_giant_2023, Park2023}, disorder can vary the chemical potential and Fermi surface, which can have drastic effects on the magnetic structure \cite{Maniv2021, Wu2022, Park2023, Xie2023, Mandujano2025}. 
%We can discuss this later as we present our experimental results. 

One member of the series, \VNS, has been proposed as a prototypical altermagnet (AM) that hosts an anomalous Hall effect (AHE) \cite{chen2021anomalous, Smejkal2022, vsmejkal2022emerging, wang_anomalous_2023, Chen2024, Ray2025, Zhu2025, Ghosh2025}. The magnetic structure of \VNS is predominantly-collinear $A$-type AFM order, where the spins within each layer are FM aligned along the $\vb{a}$-axis with neigboring layers AFM aligned \cite{Lu2020, Ray2025}. However, there have also been reports of FM \cite{Parkin1983, Inoshita2019} or $2q$ magnetic ground states \cite{Hall2021, Lancaster2023, Bentley2025, Fender2025}. In contrast to TMDs with other magnetic intercalants \cite{Mayoh2022, Park2023, Park2025}, \VNS has an anomalous Hall conductivity $\sigma_{xy}(H)$ that is not ascribed to non-collinear magnetism or spiral magnetic order, and that is much too large to be associated with the minute canted magnetic moment $M_z(H)$ through the associated Lorentz force. The large AHE instead may arise from enhanced Berry curvature \cite{Tenasini2022, wang_anomalous_2023, Zhu2025} typically associated with topological electronic band features driven by the time-reversal symmetry breaking (alter)magnetism \cite{Inoshita2019, Ghosh2025, Ray2025}. 

In this Letter, we use diffraction, neutron spectroscopy, and thermodynamic probes to determine the magnetic interactions in \VNS\ and the role of vanadium layer stacking on the physical responses of macroscopic 3D crystals. We show that there are two polytypes of \VNS with distinct vanadium layer stacking sequences that have orthogonal N\'{e}el orders. Both structures support spin wave excitations from which we obtain oscillatory exchange interactions consistent with the Ruderman-Kittel-Kasuya-Yosida (RKKY) mechanism and a giant valley-Zeeman coupling. We show that the orientation of the spontaneous magnetization and Berry curvature-driven AHE is defined by the orientation of the  N\'{e}el vector and intercalation structure. Our work thus demonstrates that \VNS is an altermagnet where electronic and magnetic degrees of freedom can be tuned by stacking engineering. 

\begin{figure} 
    \centering
    \includegraphics[width=\linewidth]{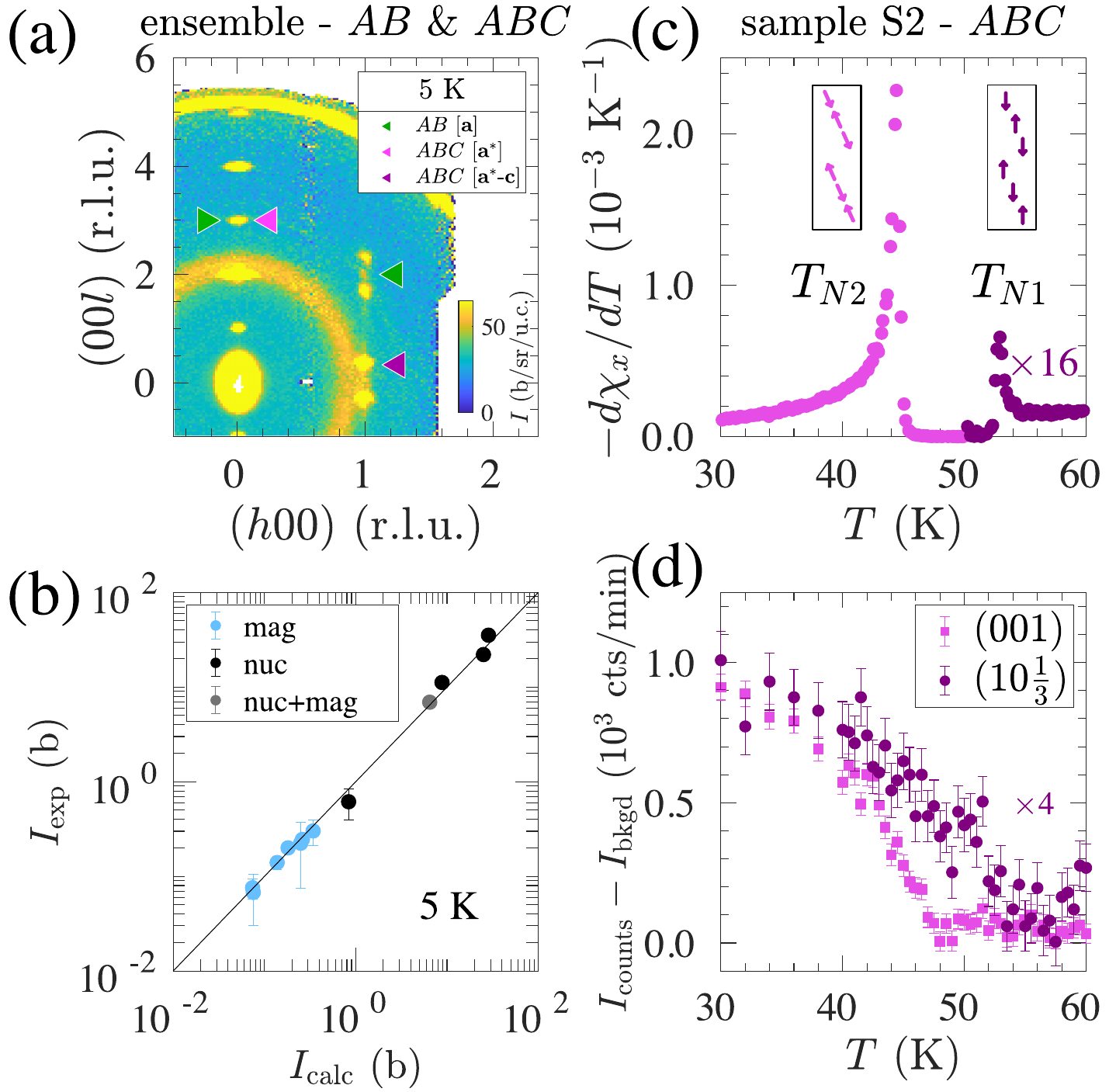}
    \caption{ \textbf{Diffraction evidence of distinct magnetic and crystal structures of \VNS.} (a) Elastic scattering in the $(h0l)$ plane at 5 K for the ensemble. (b) Simultaneous refinement of magnetic and nuclear data for the polytypes. For sample S2 with $ABC$ stacking and vanishing anomalous Hall resistivity $\Delta \rho_{xy}$, we find distinct transitions $T_N$ in the (c) differential susceptibility $(-\dv*{\chi_x}{T})$ and (d) order parameter curves on $(10\tfrac{1}{3})$ and $(001)$ magnetic peaks. These are indicative of a spin reorientation, highlighted in the inset to (c). } 
    \label{fig:Diffraction} 
\end{figure} 

%\subsection{Neutron diffraction} 
We investigate the magnetic structure and excitations using an array of co-aligned \VNS\ crystals that we denote the ``ensemble''. We also study representative single crystals (S1~and~S2) from the ensemble with different stacking types. For all crystals, we use the lattice constants of the $AB$ polytype where $a=b=5.7387$~\AA~ and $c=12.1126$ \AA\ at $T=5$~K to index the neutron and x-ray scattering data. Specifically, we represent momentum transfer by the Miller indices $(hkl)$ in reciprocal lattice units (r.l.u.) so that $\vb{Q}=h\vb{a}^* + k\vb{b}^* + l\vb{c}^*$ where $\vb{a}^*,\vb{b}^*,\vb{c}^*$ are the reciprocal lattice vectors of the corresponding hexagonal unit cell. In Fig.~\ref{fig:VNS_structure} and throughout, we work with an orthogonal Cartesian coordinate system defined so that $\vb{x} = (\bar{1}\bar{1}0)$, $\vb{y}=(1\bar{1}0)$, and $\vb{z} = (001)$ (and symmetry-related directions.) 
%Here $\vu{a}$, $\vu{b}^*$, and $\vu{c}$ can be any of the symmetry-related directions $\{100\}$, $\langle010\rangle$, and $\{001\}$, respectively.

The coherent scattering cross section of vanadium ($\sigma_{\mathrm{V}}~=~0.018$~b) is so much smaller than that of niobium ($\sigma_{\mathrm{Nb}}~=~6.25$~b) and sulfur ($\sigma_{\mathrm{S}}~=~1.02$~b), \cite{Sears1992} that neutron diffraction from \VNS is primarily sensitive to the arrangement of the $2H$-stacked \ch{NbS2} layers and not the V layers. Only below the magnetic ordering temperature $T_N\approx 50$~K do neutrons become sensitive to the vanadium stacking arrangement due to the associated magnetic order. In contrast, X-rays are sensitive to the vanadium stacking at any temperature. 

In agreement with previous reports \cite{Hall2021, Ray2025}, we observe $(00l)$ peaks for odd $l$ only for $T<T_N$ (Fig.~\ref{fig:Diffraction}(a)) in the ensemble. These are forbidden as nuclear peaks in space group P6$_3$22 and were indeed absent for $T>T_N$ in our experiment. Because neutrons are only sensitive to the magnetic moment perpendicular to wavevector transfer, the presence of magnetic $(00l)$ peaks for odd $l$ implies FM ordered spins in each basal plane \cite{SM}. We also find $(10l)$ peaks for even $l$, which confirms the $\sqrt{3}\times\sqrt{3}$ superlattice of V intercalants. Magnetic peaks at integer Miller indices indicates that the magnetic unit cell matches the chemical unit cell ($\vb{k}=\vb{0}$ order). These peaks are also observed in sample S1 which has pure $AB$ stacking (Fig.~\ref{fig:VNS_structure}(e)) and exhibits a spontaneous AHE at 2~K (Fig.~\ref{fig:SI_Hall}). 
\nocite{Sheldrick2015, Spek2020, De2010, PerezMato2015, Granroth2010, Boothroyd2020, rodriguez-carvajal_recent_1993, toth_linear_2015, Reeder2025, Frodesen1979, Scheie2022, Lygouras2024, Huberman2005, Story1992, Lee2014, blochl_projector_1994, kresse_ultrasoft_1999, Kresse_VASP, furness2020accurate, Glasbrenner2015effect, kaib2022, razpopov2023, blaha2001wien2k}

In some samples from the ensemble (such as sample S2), single-crystal X-ray diffraction (SCXRD) reveals Bragg peaks with fractional Miller $l=\tfrac{2}{3}$ (based on $c=12.1126$~\AA) for temperatures above $T_N$ (Fig.\ref{fig:VNS_structure}(a)). These indicate an alternate stacking arrangement of V layers within an otherwise identical $2H$-stacked \ch{NbS2} crystal host, as recently reported \cite{Fender2025}. This second $ABC$ polymorph belongs to the centrosymmetric space group $R\bar{3}c$ with six layers of vanadium intercalants in an $A\bar{B}C\bar{A}B\bar{C}$ sequence and lattice parameter $c'\approx 3c$ compared to the $A\bar{B}$ polytype. Here the bar indicates the inverted sulfur coordination of vanadium associated with the $2H$ NbS$_2$ structure. Evidence of the $ABC$ polytype is seen in neutron diffraction by the presence of $(10l)$ peaks for non-integer $l= \tfrac{1}{3}$ below $T_N$ (Fig.~\ref{fig:Diffraction}(a)). While there is virtually no coherent nuclear neutron diffraction from vanadium, these superlattice peaks indicate $\vb{k}=\vb{0}$ magnetic order of V$^{3+}$ moments in an $ABC$ polytype with a magnetic unit cell that, just as the chemical cell, contains $3\times 2=6$ vanadium layers. 

In a representative sample S2 (Fig.~\ref{fig:VNS_structure}(f)) with such $(10\tfrac{1}{3})$-type magnetic peaks, we find two transitions ($T_{N1}\approx 52$~K and $T_{N2}\approx 48$~K) in the magnetic susceptibility and order parameter (Fig.~\ref{fig:Diffraction}(c,d)). The absence of $(100)$ magnetic peaks in this sample implies a lack of $AB$ stacking (compare Fig.~\ref{fig:VNS_structure}(e,f)). However, whereas the $(10\tfrac{1}{3})$-type peaks emerge for $T<T_{N1}$, this sample features $(00l)$ magnetic peaks for odd $l$ that emerge for $T<T_{N2}$ (Fig.~\ref{fig:Diffraction}(d)). This implies $T_{N2}$ is a spin re-orientation transition where, in addition to the $\vb{c}$-axis component, the spins develop an in-plane AFM component upon cooling. From these data and magnetic space group theory \cite{SM}, we conclude that the $ABC$ polytype has single-$\vb{k}$, nearly-collinear AFM order with an easy axis lying in the $\vb{a}^*-\vb{c}$ plane and marginal spin canting along $\vb{a}$. In Fig.~\ref{fig:SI_Diffraction_Refinement}, we show that neither purely $\vb{a}$-axis AFM nor $120^\circ$ spin spiral order can account for the data \cite{SM}. 

%As noted in the literature \cite{Lu2020}, there is the possibility of a third structure type (denoted $AA$ stacking) consisting of the $2H$-stacked \ch{NbS2} host with partial occupancy $\approx 1/3$ on V sites. It belongs to the centrosymmetric space group $P6_3/mmc$ and does not form a $\sqrt{3}\times\sqrt{3}$ superlattice. In \cite{SM}, we show that this structure similarly has ferromagnetic in-plane oriented spins that are stacked antiferromagnetically. To confirm that the magnetism arises from distinct grains, we measured magnetic susceptibility and order parameter on a sample S1 which reveal distinct Neel temperatures (Fig.~\ref{fig:Diffraction}(c,d)).

Having provided evidence for two polytypes of \VNS with ordered $AB$ (S1) and $ABC$ stacking (S2) (Fig.~\ref{fig:VNS_structure}) we quantitatively model the ensemble of crystals used for inelastic neutron scattering as a mixture of the two polytypes. From our refinement of neutron Bragg diffraction from the ensenmble (Fig.~\ref{fig:Diffraction}(b) and Fig.~\ref{fig:SI_Diffraction_Refinement} \cite{SM}), we obtain a molar ratio $45(7):55(7)$ of $AB:ABC$ crystals, and confirm that \VNS comprises of FM layers aligned AFM (A-type AFM), with perpendicular easy axes for the two polytypes (Fig.~\ref{fig:VNS_structure}(c,d)).

\begin{figure} 
    \centering
    \includegraphics[width=\linewidth]{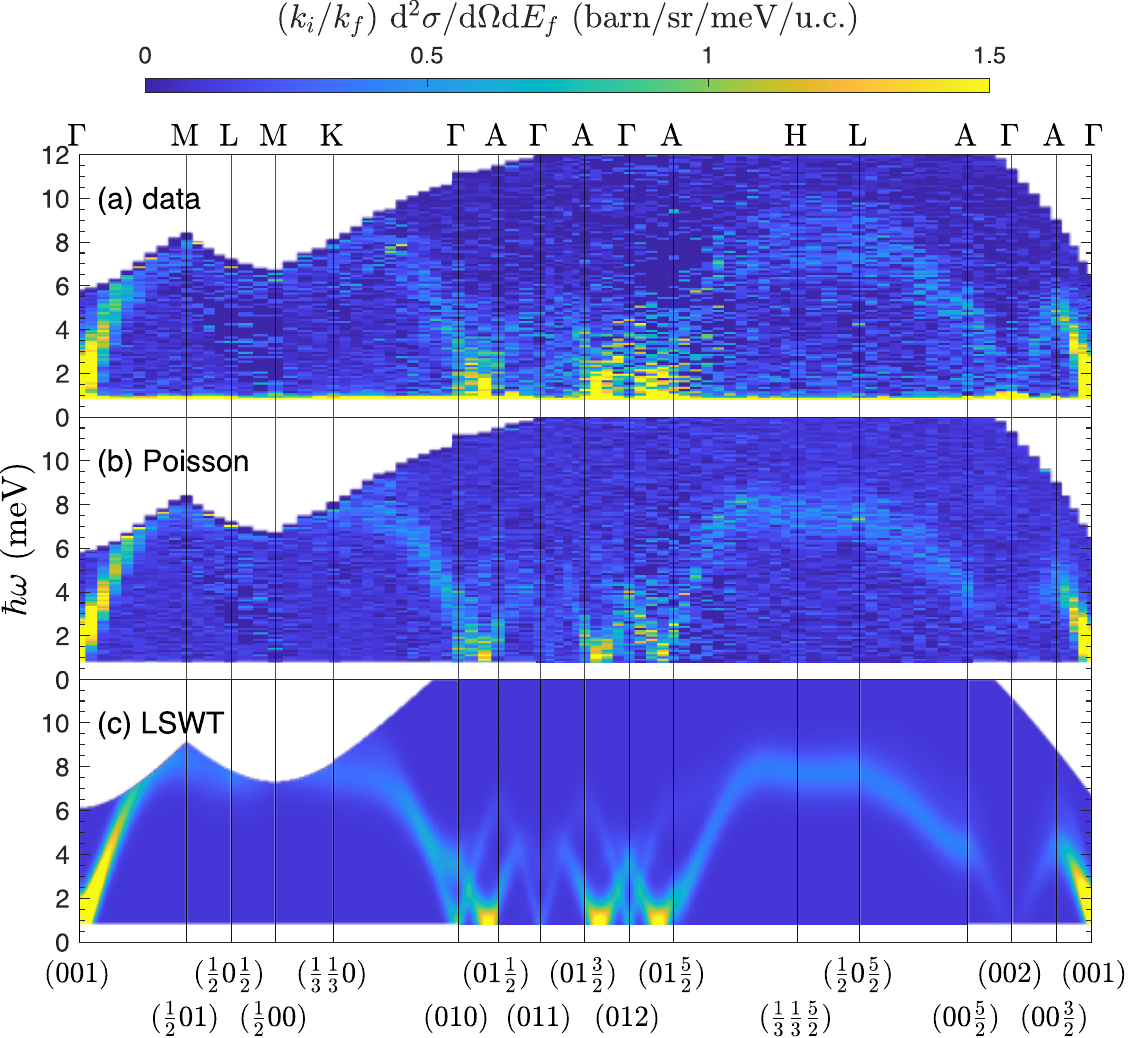}
    \caption{ \textbf{Inelastic neutron scattering from \VNS probing spin waves with wave vector transfer along high-symmetry paths in momentum space.} ``Data'' refers to the measured scattering intensity. ``Poisson'' refers to the intensity simulated considering Poisson statistics and the count rate in each pixel. ``LSWT'' refers to the theoretical intensity from linear spin wave theory simulation. Differences between spin waves in $ABC$ and $AB$ crystals are apparent  along $\Gamma(010)-A(01\frac{5}{2})$. While only high-symmetry paths are shown, the refinement was performed using an entire four-dimensional dataset.} 
    \label{fig:LSWT_Fits} 
\end{figure} 

To determine the interactions driving magnetism in \VNS, we acquired inelastic magnetic neutron scattering from the ensemble. Neutron scattering from spin waves in both polytypes produces dispersing ridges of scattering emerging from the corresponding two soft points $\vb{k}=\vb{0}$ and $\vb{k}_{\mathrm{eff}}$~=~$(00\tfrac{1}{3})$ (Fig.~\ref{fig:LSWT_Fits}(a)). At higher energies however, the data show a single magnon band, which indicates the dominant magnetic interactions are shared across polytypes. For in-plane wave vectors (along $\Gamma-K$) the bandwidth is approximately $\hslash\omega_K$~=~8~meV, while the out-of-plane bandwidth (along $\Gamma-A$) is $\hslash\omega_A$~=~4.69(5)~meV. The latter indicates substantial magnetic interactions through \ch{NbS2} layers. A two-magnon scattering peak occurring near $2\hslash\omega_K$ was recently observed with Raman spectroscopy \cite{Ghosh2025}. At the $A$-point, the magnon branch has a $1.6(2)$~meV full width at half-maximum (FWHM) compared to the instrumental resolution of 1.0~meV. The remaining physical width of 1.2(2)~meV provides a measure of differences in exchange parameters between polytypes. Disorder associated with the coexistence of polytypes in a given crystal could also contribute to the enhanced linewidth. The physical width also constrains the possible size of altermagnetic magnon splitting in this experiment on an unmagnetized sample in our unpolarized neutron experiment \cite{Morano2025, Faure2025, Singh2025}.
%, and the instrument resolution constrains the maximum linewidth for quasielastic scattering from heavy quasiparticle fluctuations \cite{Regnault1987}. 

For a quantitative measure of V-V exchange interactions we fit these data to $1/S$ spin wave theory based on the following Heisenberg model
 \begin{equation} \label{eq:Hamiltonian}
     \mathcal{H} = \sum_{ij} J_{ij} \vb{S}_{i} \cdot \vb{S}_{j} - \sum_{i} \vb{S}_i^T D_{\vu{n}} \vb{S}_i .
 \end{equation} 
Here $D_{\vu{n}}$ is a phenomenological magnetocrystalline anisotropy tensor with easy axis $\vu{n}$. Due to the correspondence in the V-V coordination among the two polytypes up to the sixth near-neighbor (Fig.~\ref{fig:ExchangeBonds}), we approximate the exchange interactions $J_1-J_7$ as isotropic and identical across polytypes and only allow for the direction of the easy axis defined by $D_{\vu{n}}$ to differ in accordance with the ordered spin structure. The addition of $J_7$ optimizes the goodness of fit, and improves the prediction of the critical and Curie-Weiss temperatures \cite{SM}. The single-ion anisotropy $|D_{\vu{n}}|=0.03(5)$~meV, constrained by the lack of a gap in the measured excitation spectrum, is consistent with the value of $0.05(1)$~meV obtained from density functional theory (DFT) \cite{SM,riedl2022}. 
Our analysis is insensitive to exchange anisotropy or Dzyaloshinskii–Moriya interactions that may also play a role in stabilizing the different spin structures \cite{Yang2023}. We extract the exchange parameters $J_n$ using a pixel-to-pixel fit of the scattering intensity over the entire four-dimensional dataset \cite{SM}. The similarity of the measurement (Fig.~\ref{fig:LSWT_Fits}(a)) to the simulated data (Fig.~\ref{fig:LSWT_Fits}(b)) provides confidence in the inferred interaction parameters. Fig.~\ref{fig:LSWT_Fits}(c) shows the theoretical intensity distribution, which shows that contributions from the two polytypes are distinguishable only near the critical wave vectors. 

\begin{figure}
    \centering
    \includegraphics[width=\linewidth]{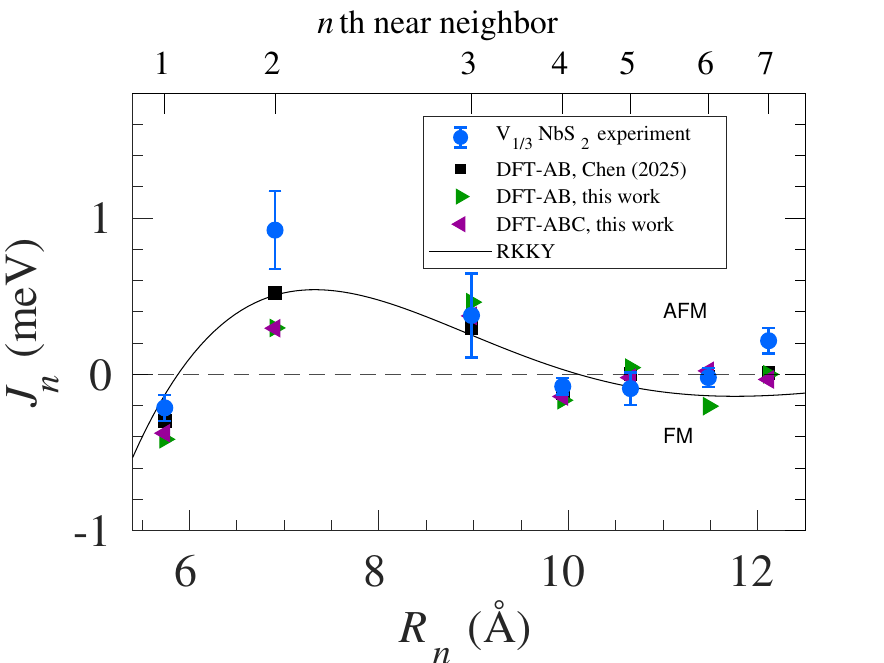}
    \caption{ \textbf{Heisenberg exchange constants $J_n$ of \VNS as a function of near-neighbor distance $R_n$.} The dotted line shows the fit to Eq~\ref{Eq:RKKY}. We include the DFT-predicted values from Chen \textit{et al.} \cite{Chen2024}, and this work. Error bars represent one standard deviation. } 
    \label{fig:ExchangeFit} 
\end{figure} 

The exchange constants $J_n$ thus inferred are shown as a function of the near-neighbor distances between spins, $R_n$, in Fig.~\ref{fig:ExchangeFit} and Table~\ref{tab:J_couplings_DFT} with comparison to DFT results by Chen \textit{et al.}~\cite{Chen2024} and obtained in this work~\cite{SM}. The experiment and DFT find oscillatory interactions extending through 1 nm with FM nearest-neighbor ($J_1<0$) and AFM next-nearest neighbor $J_2>0$ that stabilize $A$-type AFM \cite{Ray2025, Ghosh2025}. The Luttinger-Tisza method \cite{friedman_solution_1974} shows that, over a wide range of parameter space, the $\vb{k}=\vb{0}$ AFM order is stable with respect to changes in exchange parameters that could be induced by chemical disorder, doping, or polymorphism (Fig.~\ref{fig:SI_LuttingerTisza} in \cite{SM}).

RKKY exchange interactions mediated by a spherical Fermi surface take the following oscillatory form (Eq.\ref{eq:RKKY_Amplitude}): 
\begin{equation}
     J(R)= \frac{9\pi}{8^2} \frac{J_K^2}{E_F} \left(\frac{2k_F R\cos(2k_F R)-\sin(2k_F R)}{(k_F R)^4} \right) 
     \label{Eq:RKKY}
\end{equation} 
where $J_K$ is the Kondo interaction strength to conduction electrons with band mass $m^*$ and Fermi wave vector $k_F$. A good fit (Fig.~\ref{fig:ExchangeFit}) is obtained for $k_F = 0.382(3)$ \AA$^{-1}$. Previous investigations \cite{Lancaster2023, edwards_giant_2023, Ray2025} identified ellipsoidal Fermi surface pockets with an estimated $k_F\approx 0.2-0.45$~\AA$^{-1}$. In our DFT calculations, the closest matching wavevector nests nearly-flat regions of the Fermi surface (Fig.~\ref{fig:VNS_nesting}). Note that $J(R)$ passes from FM to AFM interactions for $R=4.4934/2k_F=5.8~$\AA, which  lies between $R_1$ and $R_2$ as expected for the $A$-type AFM. The overall magnitude of the RKKY interaction \cite{Ruderman1954, SM} provides an estimate of $J_K = 91(7)$~meV for the Kondo coupling. 

%findings from neutron scattering 
%\VNS is part of a growing list of metals and semimetals where direct evidence for oscillatory RKKY interactions has been obtained from neutron spectroscopy \cite{Scheie2022, Lygouras2024, Boothroyd2025}. 
Our fit to the magnon spectrum provides evidence for oscillatory RKKY interactions in \VNS with effective Fermi wavenumber $k_F \approx 0.382(3)$~\AA$^{-1}$ and Kondo coupling strength $J_K= 91(7)$~meV. The latter is comparable to the value obtained from ARPES by Edwards \textit{et al.} \cite{edwards_giant_2023} who estimated an interaction scale of $50$~meV between V moments and conduction electrons from \ch{NbS2}. They proposed that the uncompensated $\vb{c}$-oriented spin components beneath the surface layer of \ch{NbS2} drive a giant internal Zeeman field $J_K S\sim 250~\mathrm{T}$, and the coupling to conduction electrons leads to a valley-selective electronic band splitting $J_K \pm \Delta_{\mathrm{SOC}}$ at $K$ and $K'$ respectively, where $\Delta_{\mathrm{SOC}}$ is the spin-orbit coupling gap. Such effect is naturally explained if the samples have $ABC$ stacking with $\vb{c}$-axis component of spins, as determined in our work. Alternatively, it could arise in purely $AB$-stacked crystals with bulk in-plane AFM if the anisotropy energy favors out-of-plane spin reorientation at the surface.

\begin{table}
\centering \footnotesize 
\caption{V-V exchange interactions in \VNS from fitting to experiment and calculations within DFT  for $S~=~1$ spins. For each bond we also write the coordination numbers $z_n$, relative spin orientation $\sigma_n=S_n/S_1$ for the spin structure, and average bond separation distance $\bar{R}_n$. Unfrustrated interactions satisfy $\sigma_n = -\mathrm{sgn}(J_n)$. Following the convention in Eq.~\ref{eq:Hamiltonian}, negative (positive) interactions are FM (AFM). The DFT values in this work are calculated using meta-GGA.} 
\begin{tabular}{cccc|cccc}
    $J_n$ & $z_n$ & $\sigma_n$ & $\bar{R}_n$ (\AA) & \textbf{Exp.} & \textbf{DFT} & \textbf{DFT - AB} & \textbf{DFT - ABC} \\
    (meV) &       &            &       & (this work) & \cite{Chen2024} & (this work) & (this work) \\
    \hline
    $J_1$ & 6      & $+1$      & 5.74  & $-0.21(8)$   & $-0.298$ & $-0.416$ & $-0.377 $\\        %& $-0.709$ & $-0.342$ \\
    $J_2$ & 6      & $-1$      & 6.98  & $+0.9(2)$   & $+0.519$ & $+0.297$ & $+0.295$ \\         % & & $+0.379$ $+0.210$ \\
    $J_3$ & 6      & $-1$      & 9.04  & $+0.4(3)$   & $+0.297$ & $+0.462$ & $+0.374$ \\      %  & $+0.297$ & $+0.275$ \\
    $J_4$ & 6      & $+1$      & 9.95  & $-0.08(5)$  & $-0.118$ & $-0.166$ & $-0.142$ \\         % & $-0.087$  & $-0.157$ \\
    $J_5$ & 12     & $-1$      & 10.71 & $-0.1(1)$ & $+0.001$ & $+0.043$ & $-0.021$ \\       %  &  $+0.036$ & $+0.071$\\
    $J_6$ & 6      & $+1$      & 11.49 & $-0.01(6)$  & $-0.013$ & $-0.204$ & $+0.022$ \\      % & $-0.029$  & $+0.008$ \\
    $J_7$ & $2$~|~$6$ & $+1$   & 12.5 & $+0.21(8)$  & $+0.010$ & $-0.000$     &  $-0.033$ \\       % & $-$   & $-$ \\
\end{tabular}
\label{tab:J_couplings_DFT}
\end{table}

\begin{figure} 
    \centering
    \includegraphics[width=\linewidth]{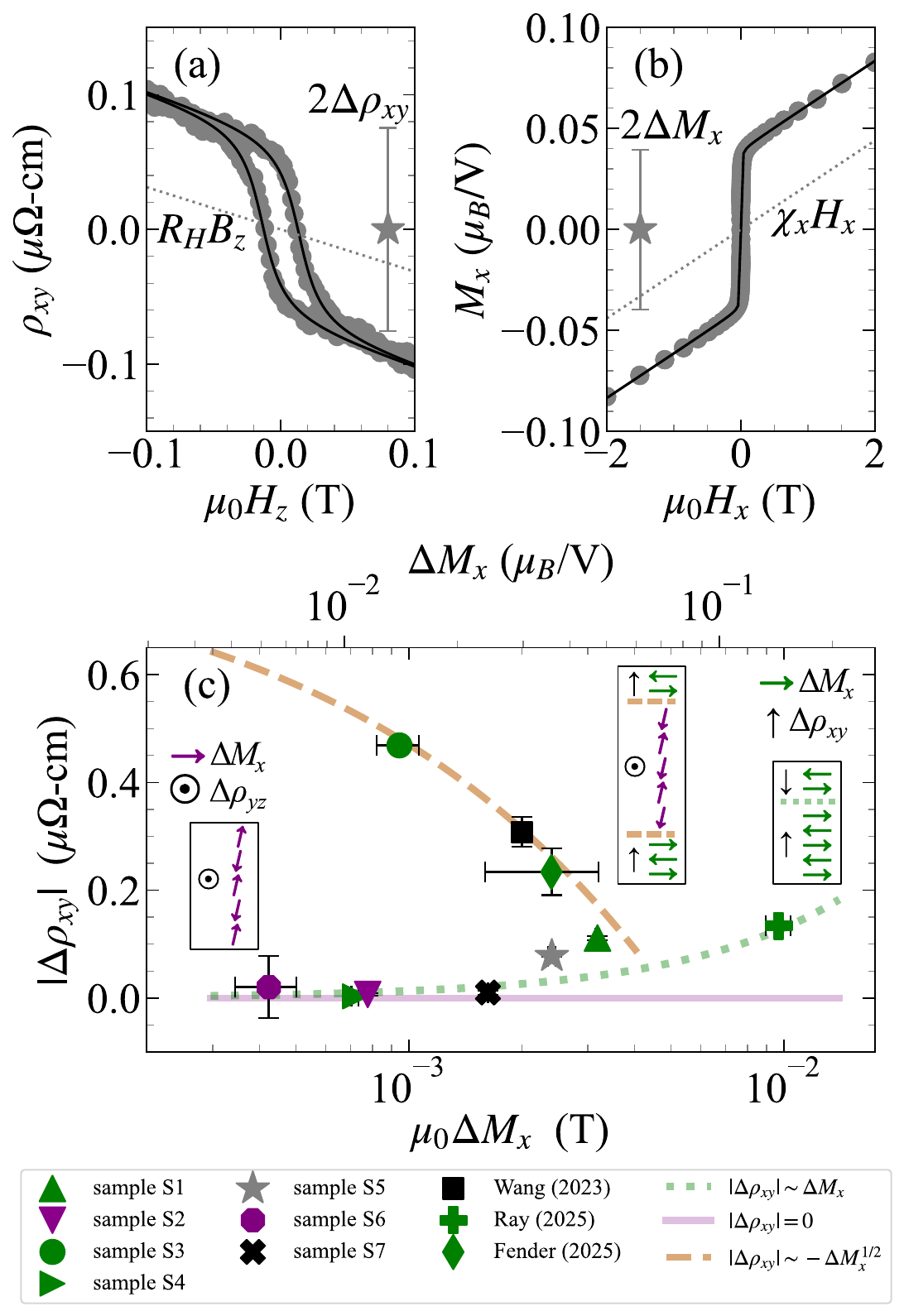} 
    \caption{ \textbf{In-plane uncompensated magnetism versus out-of-plane anomalous Hall resistivity in \VNS.} (a) Hall resistivity $\rho_{xy}$ and (b) in-plane magnetization $M_x$ for sample S5 as a function of magnetic field at 2 K with fits described in Eq.~(\ref{eq:TwoFunctionFits}). (c) Semi-logarithmic plot of $|\Delta \rho_{xy}|$ versus in-plane remanent magnetization $\mu_0 \Delta M_x$ across various samples at 2 K. 
    The insets show how structural domains with odd number of spins support extrinsic in-plane FM, with the local Hall vectors in black arrows, and the resulting functional dependence of AHE shown by the colored lines. Hall resistance data for samples S1-S6 are shown in Fig.~\ref{fig:SI_Hall}. When explicitly measured with diffraction, green (purple) markers refer to $AB$ ($ABC$) polytypes.} 
    \label{fig:UniversalHall} 
\end{figure}

%stacking and AHE 
A confounding feature of \VNS\ has been that magnetization and transport properties of nominally stoichiometric crystals can display considerable variability even from a single batch. We find that the distinct magnetic properties of coexisting polytypes provide an explanation. %This may be understood as a consequence of the admixture of mesoscopic slabs of the two polytypes {\color{red} (also known as allotwins). The occurrence of such structures has been observed in electron microscopy \cite{Fender2025}}, and are apparent in the partial occupancy of V sites that is necessary to describe SCXRD from some samples. 
For A-type AFM, lamellar domains with an odd number of vanadium planes results in a net FM moment oriented along the N\'{e}el vector (see insets of Fig.~\ref{fig:UniversalHall}(c) and \cite{SM}). Fig.~\ref{fig:UniversalHall}(a,b) show that a finite remnant in-plane magnetization $\Delta M_x(H \to 0)$ indeed can coexist with out-of-plane anomalous Hall resistivity $\Delta \rho_{xy}(H\to0)$. 

Fig.~\ref{fig:UniversalHall}(c) shows $\Delta \rho_{xy}(H\to0)$ versus $\Delta M_x(H \to 0)$ for nominally stoichiometric \VNS\ grown under a variety of conditions and in various laboratories \cite{wang_anomalous_2023, Ray2025}. The bifurcation of the data is indicative of distinct allotwinned samples. 
For predominantly $ABC$-stacked samples, $\Delta M_x(H \to 0)$ arises from in-plane spin canting or odd number of layers, while $\Delta\rho_{xy}=0$ by symmetry (see left inset and solid purple line in Fig.~\ref{fig:UniversalHall}(c)). Thus, we associate the lower left branch in Fig.~\ref{fig:UniversalHall}(c) with predominantly $ABC$-stacked samples. Single-domain $AB$ samples are compensated AFM so that $\Delta M_x\to0$ while $\Delta \rho_{xy}$ is maximal corresponding to the upper-left branch in Fig.~\ref{fig:UniversalHall}(c). The nonzero $\Delta M_x(H \to 0)$ in predominantly $AB$-stacked samples is an extrinsic effect arising from domains with an odd number vanadium layers (see right insets in Fig.~\ref{fig:UniversalHall}(c)).

We posit two regimes of extrinsic defect-induced domains might influence otherwise intrinsic AHE. The dashed and dotted lines in Fig.~\ref{fig:UniversalHall}(c) compare the data to models of defect-induced domains in the two types of samples. In primarily $AB$-stacked crystals where domains arise from FM-coupled $AB-BA$ stacking faults (right-most inset of Fig.~\ref{fig:UniversalHall}(c)), the net in-plane magnetization is enhanced while the net Hall vector decreases, resulting in a functional dependence $\Delta \rho_{xy}\sim\Delta M_x$ as shown by the dotted green line. This results in some $AB$ samples having nearly-vanishing (e.g.~sample S4) or enhanced (e.g.~from Ray \textit{et al.} \cite{Ray2025}) remanent in-plane magnetization and out-of-plane AHE.
%Some $AB$ samples have enhanced Hall effect with increasing in-plane magnetization, and we associate these (in the lower right branch) as having FM domain walls.} 
Meanwhile, when domains involve $ABC$ stacks introduced between $AB$ regions (center-right inset of Fig.~\ref{fig:UniversalHall}(c)), the Hall vector similarly decreases, and it can be shown that $\Delta \rho_{xy}\sim -\Delta M_x^{1/2}$\cite{SM} as indicated by the dashed brown line. The good qualitative account of the data in Fig.~\ref{fig:UniversalHall}(c) suggests these to be plausible interpretations. Furthermore, both models counterintuitively predict that $\Delta M_x$ should increase in thinner samples that have a smaller number of defects and layers -- a prediction that we verified by repeated measurements on a single sample (S7 in Fig.~\ref{fig:UniversalHall}(c)) after thinning through mechanical polishing (see Fig.~\ref{fig:VNS_polishing}). 

%altermagnetism 
\VNS lies within a broad class of altermagnets \cite{Smejkal2020, vsmejkal2022emerging, Cheong2025} with a significant AHE despite compensated collinear AFM order and vanishingly small spin canting. While both polytypes exhibit A-type AFM, the spin directions change from $\vb{a}$-direction for $AB$-stacked crystals to the $\vb{a}^*-\vb{c}$ plane for $ABC$-stacked crystals. For the $ABC$ polytype, the spins lie nearly in the plane of the sulfur octahedra, which suggests that single ion crystal field effects play a significant role. In \cite{SM}, we detail the symmetry-allowed Hall tensors associated with these magnetic structures. We find that the spin structure of $AB$-stacked ($ABC$-stacked) crystals allows for nonzero anomalous Hall conductivity $\sigma_{xy}$ $(\sigma_{yz}$) and slight spin canting along $\vu{z}$ ($\vu{x})$; see Fig.~\ref{fig:VNS_structure}(c,d). Importantly, both structures are consistent with altermagnetic ground states. This is in line with our DFT calculations which reveal zero net magnetization and altermagnetic spin splitting in the electronic band structure along certain high-symmetry directions \cite{SM}.

%Hall effect 
%%Importantly, we confirm that $2q$ non-collinear magnetism is not responsible for AHE \cite{Ray2025, wang_anomalous_2023, Hall2021}. 
It was previously established that \VNS lies in a regime of large intrinsic AHE in $\sigma_{xy}$ due to the enhanced Berry curvature from band topology. Indeed, Weyl nodes were predicted and experimentally verified in \VNS \cite{Inoshita2019, Ghosh2025}. These findings contrast with the scaling $\sigma_{yz} \propto M_x$ for conventional FM \cite{Nakatsuji2015, Zeng2006} found in some regimes of \VNS \cite{Ray2025}. Our work suggests that $\sigma_{yz}$ arises from $ABC$ grains which feature slight in-plane spin canting, or from $AB$ slabs with an odd number of layers separated by $ABC$ stacks. Importantly, the $AB$ polytype is noncentrosymmetric and thus can host Weyl fermions, while the $ABC$ structure is centrosymmetric and cannot accommodate Weyl fermions unless through time-reversal symmetry breaking. This could explain the different origins of the Hall effect: $\sigma_{yz}$ arises from spin-canting AFM, whereas $\sigma_{xy}$ arises from Berry curvature.

The physical properties of bulk \VNS mirror phenomena previously found in the atomic limits of vdW magnets. The remanent in-plane magnetization in this compensated AFM occurs from domains with an odd number of spins. This is akin to the net FM arising in atomically-thin odd-layer AFMs, contrasting with the even-layer case in which the net moment is zero \cite{Fei2018, Yang2021}. Additionally, layer stacking affects local properties such as the easy axis of magnetization, and this effect likewise occurs in some vdW magnets toward the monolayer limit \cite{Yang2023, Hanke2017}. Together, these local properties either promote or impede enhanced Berry curvature and AHE in \VNS. This motivates the tantalizing prospect of realizing new magnetic ground states, charge density waves, or superconductivity by controlling layer stacking in the bulk limit. %Conversely, it may be of interest to study the effect of disparate easy-axis AM or uncompensated FM on anomalous electron transport toward the atomic limit. 

In summary, \VNS supports anisotropic, carrier-mediated $A$-type AFM in distinct polytypes. The interaction strengths and altermagnetic order are robust to changes in stacking or polytypism.
Our results show that it may be possible to use intercalation staging to program magnetic and magnetotransport properties in van der Waals solids.

\section{Contributions}
The crystals were assembled and measured on SEQUOIA by C.J.L., N.P., and Y.C. under the supervision of A.I.K., R.J.B. and C.L.B. The inelastic neutron scattering data were analyzed and fit by C.J.L. and J.H.D. Representative crystals were measured on DEMAND and TOPAZ by C.J.L. and Y.C. under the supervision of C.H., Y.H. and H.C. The SCXRD measurements and refinements were performed by M.A.S. The synthesis, magnetization, and transport measurements were performed by Z.F., M.F., L.F. and C.J.L. under the supervision of S.N. and C.L.B. The DFT calculations were performed by L.G.G. under the supervision of R.V. 

\section{Acknowledgments} 
We would like to thank Predrag Nikolic, Andrei Savici, Takumi Matsuo and Timothy Reeder for valuable discussions. 
This work was supported by the Department of Energy, Office of Science, Basic Energy Sciences under Award No. DE-SC0024469. C.B. was supported by the Gordon and Betty Moore Foundation EPIQS program under GBMF9456. The Institute for Quantum Matter was supported through a generous donation by William H. Miller III to the Department of Physics and Astronomy. This work was partially supported by JST-Mirai Program (JPMJMI20A1), JST-ASPIRE(JPMJAP2317). %updated as at 20250916
Work at the University of California, Berkeley and Lawrence Berkeley National Laboratory was funded by the U.S. DOE, Office of Science, Office of Basic Energy Sciences, Materials Sciences and Engineering Division under Contract No. DE-AC02-05CH11231 (Quantum Materials Program KC2202).
Computational analysis was carried out at the Advanced Research Computing at Hopkins (ARCH) core facility  (rockfish.jhu.edu), which is supported by the National Science Foundation (NSF) grant number OAC1920103.
A portion of this research used resources at the Spallation Neutron Source, and the High Flux Isotope Reactor, DOE Office of Science User Facilities operated by the Oak Ridge National Laboratory. 
Beam time was allocated to the SEQUOIA spectrometer on proposal number IPTS-27225.1; to the DEMAND diffractometer on IPTS-31587.1; and to the TOPAZ diffractometer on IPTS-34423.1. Part of the research of supported by the Deutsche Forschungsgemeinschaft (DFG, German Research Foundation) through the TRR 288 - 422213477. 
This manuscript has been authored in-part by UT-Battelle, LLC, under contract DE-AC05-00OR22725 with the US Department of Energy (DOE). The US government retains and the publisher, by accepting the article for publication, acknowledges that the US government retains a nonexclusive, paid-up, irrevocable, worldwide license to publish or reproduce the published form of this manuscript, or allow others to do so, for US government purposes. DOE will provide public access to these results of federally sponsored research in accordance with the DOE Public Access Plan ( \url{https://www.energy.gov/doe-public-access-plan} ).

\bibliography{bibliography}

% ---------- SUPPLEMENT ----------
\clearpage
\setcounter{figure}{0}
\renewcommand{\thefigure}{S\arabic{figure}}

\setcounter{figure}{0}
\counterwithin*{figure}{section}
\renewcommand{\thefigure}{S\arabic{figure}}

%restart figure and table numbering 
%\setcounter{figure}{0} 
\setcounter{equation}{0}  
\setcounter{table}{0} 
\renewcommand{\theequation}{S\arabic{equation}}
\renewcommand{\thetable}{S\Roman{table}}

\section{Supplementary Material (SM)}

\subsection{Single-crystal X-ray diffraction (SCXRD)}
All reflection intensities were measured at 213(2) K, well above the magnetic ordering temperature, using a SuperNova diffractometer (equipped with Atlas detector) with Mo K$\alpha$ radiation ($\lambda = 0.71073$~\AA) under the program CrysAlisPro (Version CrysAlisPro 1.171.42.49, Rigaku OD, 2022). The same program was used to refine the cell dimensions and for data reduction. The structure was solved with the program SHELXS-2018/2 and was refined on $F^2$ with SHELXL-2019/3. Empirical absorption correction using spherical harmonics was applied using CrysAlisPro. The temperature of the data collection was controlled using the Cryojet system manufactured by Oxford Instruments. \cite{Sheldrick2015}. 

Since there has been no evidence of charge density wave order in the thermodynamic probes of \VNS, we do not anticipate superlattice peaks beyond those arising from the intercalated $\sqrt{3}\times\sqrt{3}$ structure. Hence, the non-integral $l=2/3$-type peaks observed in the XRD precession images in some samples indicate a distinct new polytype (Fig.~\ref{fig:VNS_structure}(a)). Given that in the \ch{NbS2} layers above and below each intercalant layer are rotated relative to each other by 180$^\circ$ in $2H$-stacking, the $ABC$ polytype must have $c'\approx 3c$ to form a Bravais lattice. 

During the CHECKCIF analysis, ADDSYM/PLATON \cite{Spek2020} strongly suggested the space group $P6_3/mcm$ for the AB polytype; however, refinement in this space group yields poor results. Furthermore, numerous systematic absence violations indicate that $P6_3/mcm$ is likely incorrect. 

\subsection{Magnetic space group symmetry}
In the following, we describe the important details of the symmetry analysis for the two polytypes and the anticipated Hall effects. We use the concept of the Hall vector $\vb{h}$ for which the anomalous Hall current $\vb{j}_H = \vb{h}\times\vb{E}$ where $\vb{h} = (\sigma_{zy}, \sigma_{xz}, \sigma_{yx})$. A nonzero Hall vector indicates the components of the Hall conductivity that are allowed by symmetry. 

In the magnetic space group approach \cite{De2010,Cheong2025, PerezMato2015}, the structure of $AB$ stacked \VNS belongs to the magnetic space group $C2'2'2_1$ (\#20.33) with associated magnetic point group $2'2'2$, as described in \cite{Ray2025}. This group allows for a Hall vector $\vb{h}$ aligned along the rotation axis, i.e., along the $z$-axis. Hence, the magnetic point group only allows for a Hall effect $\sigma_{xy}$. Note that this holds strictly for $\vb{a}$-axis oriented spins. If the spins were oriented along the $\vb{a}^*$-axis, the Hall vector would instead vanish. 

%Meanwhile, the structure of disordered $AA$ layer stacking and local $\vb{a}$-axis moments (in the lattice units of the space group $P6_3/mmc$) is consistent with the magnetic point group $mmm$, which is centrosymmetric and incompatible with ferromagnetism, and thus cannot accommodate an anomalous Hall effect. It is interesting to note that in the lattice units of the space group $P6_322$, with a $\pi/3$ rotation between the coordinate systems, the spins appear oriented along the $\vb{a}^*$ axis. 

Next, \VNS with $ABC$-stacking has a magnetic structure consisting of collinear AFM spins. We first note the convention of the orthogonal coordinate system where $\vu{x}=\vu{a}$ and $\vu{z}=\vu{c}$. There are two magnetic phase transitions in $ABC$ crystals. For $T_{N2}<T<T_{N1}$, the $(001)$ magnetic peak is absent while $(10\tfrac{1}{3})$ onsets. This implies the staggered magnetization is oriented along $\bf c$ so that magnetic diffraction at $(001)$ is extinguished by the polarization factor in the magnetic neutron scattering cross section. This first spin structure, with magnetic space group $R\bar{3}c$ (\#167.103), has an associated magnetic point group $\bar{3}m$. These retain all spatial-symmetry elements, combined with translation, but break time-reversal symmetry. Below the second phase transition $T<T_{N2}$, an in-plane spin component onsets as evidenced by the appearance of intensity at $(001)$. The latter structure is consistent with the magnetic space group $C2/c$ (\#15.85), having inversion, and nonsymmorphic rotation $\{2_{100}|(0,0,1/2)\}$ and mirror $\{ m_{100} | (0,0,1/2)\}$ symmetries. The reduction of additional spatial symmetry elements is consistent with the experimental indications of a phase transition and not a crossover. These symmetry operations of $C2/c$ allow for a FM component along $\vu{x}=\vu{a}$ and a staggered component in the perpendicular $\vu{y}-\vu{z}=\vu{a}^*-\vu{c}$ plane. The magnetic point group associated with the spin structure is $2/m1$ (\#5.1.12) with a twofold rotation axis $\vu{a}=(100)$. This allows for a Hall vector along the rotation axis $\vu{a}=\vu{x}$ and thus a finite Hall conductivity $\sigma_{yz}$.

\iffalse 
\begin{table*} 
\centering
\begin{tabular}{c|c|c|c|c} 
\toprule
Polytype & Mag. point group & Elements & Phase & FM? \\ 
\hline 
AB & $622.1'$ & $1, 6^+_{001}, 3^+_{001}, 2_{001}, 3^-_{001}, 6^-_{001}, 2_{100}, 2_{210}, 2_{110}, 2_{120}, 2_{010}, 2_{1-10},$ & $T>T_N$ & No \\
  &  &  $1', 6'^+_{001}, 3'^+_{001}, 2'_{001}, 3'^-_{001}, 6'^-_{001}, 2'_{100}, 2'_{210}, 2'_{110}, 2'_{120}, 2'_{010}, 2'_{1-10}$ & &  \\ \hline 
   & $2’2’2$ & $1, 2_{001}, 2’_{100}, 2’_{010}$ & $T<T_N$ & Yes \\ \hline \hline 
ABC & $\bar{3}m.1'$ & $1, 3^+_{001}, 3^-_{001}, 2_{100}, 2_{110}, 2_{010}, -1, -3^+_{001}, -3^-_{001}, m_{100}, m_{110}, m_{010}$, & $T>T_{N1}$ & No \\ 
  &  & $1', 3'^+_{001}, 3'^-_{001}, 2'_{100}, 2'_{110}, 2'_{010}, -1', -3'^+_{001}, -3'^-_{001}, m'_{100}, m'_{110}, m'_{010}$ &  &  \\ \hline 
 & $\bar{3}m.1$ & $1, 3^+_{001}, 3^-_{001}, 2_{100}, 2_{110}, 2_{010}, -1, -3^+_{001}, -3^-_{001}, m_{100}, m_{110}, m_{010}$ & $T_{N2}<T<T_{N1}$ & No \\ \hline 
 & $2/m.1$ & $1, 2_{010}, -1, m_{010}$ & $T<T_{N2}$ & Yes \\ \hline 
\hline 
\end{tabular}
\caption{Summary of the magnetic point groups for the $AB$ and $ABC$ polytypes. } 
\label{tab:Symmetry_table}
\end{table*}
\fi 

Table~\ref{tab:VNS_table} highlights the different space groups, spin alignments, possible nonzero AHE tensors, and selection rules for select magnetic peaks. 

\begin{table} 
\centering
\begin{tabular}{c|c|c|c} 
\toprule
Stacking & $AB$ & $ABC$ & $ABC$ \\ 
\hline 
Space group & $P6_322$ & $R\bar{3}c$ & $R\bar{3}c$ \\ 
Spin component &  $\vb{a}$-axis & $\vb{a}^*$-axis & $\vb{c}$-axis \\ 
Mag. point group & $2'2'2$ & $2/m$ & $2/m$ \\
AHE & $\sigma_{xy}$ & $\sigma_{yz}$ & $\sigma_{yz}$ \\ 
$(100)$? & $\checkmark$ & $\cross$ & $\cross$  \\ 
$(001)$? & $\checkmark$ & $\checkmark$ & $\cross$ \\ 
$(10\tfrac{1}{3})$? & $\cross$ & $\checkmark$ & $\checkmark$ \\ 
\hline 
\end{tabular}
\caption{Summary of the crystal types possible for \VNS, with its associated magnetic structures and Hall effect. The coordinate system refers to the $\sqrt{3}\times\sqrt{3}$ ordered structures.} 
\label{tab:VNS_table}
\end{table}

%insert table for point groups associated with the various magnetic phases in the various polytypes  

\subsection{Neutron diffraction} 
Measurements of the multi-crystal ensemble were done on the SEQUOIA time-of-flight spectrometer at Oak Ridge National Laboratory \cite{Granroth2010}. Diffraction experiments on single crystals S1 and S2 were done on the DEMAND and TOPAZ neutron diffractometers. 

The differential neutron scattering cross-section for elastic diffraction from ordered moments is \cite{Boothroyd2020, rodriguez-carvajal_recent_1993} 
 \begin{eqnarray}
     \dv{\sigma}{\Omega} &=& N \left( \frac{\gamma r_0}{2\mu_B}\right)^2 \frac{(2 \pi)^3}{v} |f(Q)|^2 \\ \nonumber
     &\times&\sum_{\alpha\beta} \left( \delta_{\alpha\beta} - \frac{Q_\alpha Q_\beta}{Q^2} \right) \sum_{\vb*{\tau}} (F_M^\alpha(\vb{Q}))^* (F_M^\beta(\vb{Q})) \delta(\vb*{\tau} - \vb{Q}),
 \end{eqnarray} 
where $\gamma= -1.913$ is the magnetic moment of the neutron in units of the nuclear Bohr magneton, $r_0=(1/4\pi\epsilon_0)(e^2/m_ec^2)= 2.82$~fm is the classical electron radius, $f(Q)$ is the magnetic form factor, and the summation is over cartesian components. The vector magnetic structure factor is given by
 \begin{equation} 
     \vb{F}_M(\vb{Q}) = \vb{F}_M (\vb{H}+\vb{q}) = \sum_{\vb{d}} \vb{m}_{\vb{d}}(\vb{q}) e^{i\vb{H}\cdot \vb{d}}, 
     \label{FmQ}
 \end{equation} 
with $\vb{H}$ a reciprocal lattice vector, $\vb{q}=\vb{0}$ is the magnetic ordering wavevector, and $\vb{m}_{\vb{d}} = \vb{m}_{\vb{d}}(\vb{0})$ is the magnetic dipole moment on the site with position vector $\vb{d}$. The three-dimensional momentum space coverage allows us to determine the nuclear and magnetic structure factors by integrating the scattering data around each Bragg peak employing $\int d^3\vb{q}\delta(\vb{q})=1$. 

Throughout, we consider the reciprocal lattice units of the ordered $\sqrt{3}\times\sqrt{3}$ structure consistent with the $AB$ polytype. In Fig.~\ref{fig:Diffraction}(d,e), we show neutron diffraction data from TOPAZ in the $(h0l)$ plane at 5~K for two distinct crystals. Both samples were characterized by x-ray diffraction. In the 0.1 mg sample S1, which does have AHE ($\sigma_{xy}\ne 0$), we only find evidence of the $(003), (100)$, and $(102)$ magnetic peaks. In sample S2 (without AHE), we observe both non-integral $(10\tfrac{1}{3})$ type peaks of the ABC stacking along with $(00l)$ peaks for odd $l$. However, this particular sample lacks $(100)$ peaks, which implies that it does not have AB grains. From this we infer that the apparent $\vb{k}_{\mathrm{eff}} = (00\tfrac{1}{3})$ order seen in the multi-crystal sample arises from $ABC$-stacked samples within the ensemble and is absent in AB polytype samples. 

\begin{figure}
    \centering
    \includegraphics[width=0.825\linewidth]{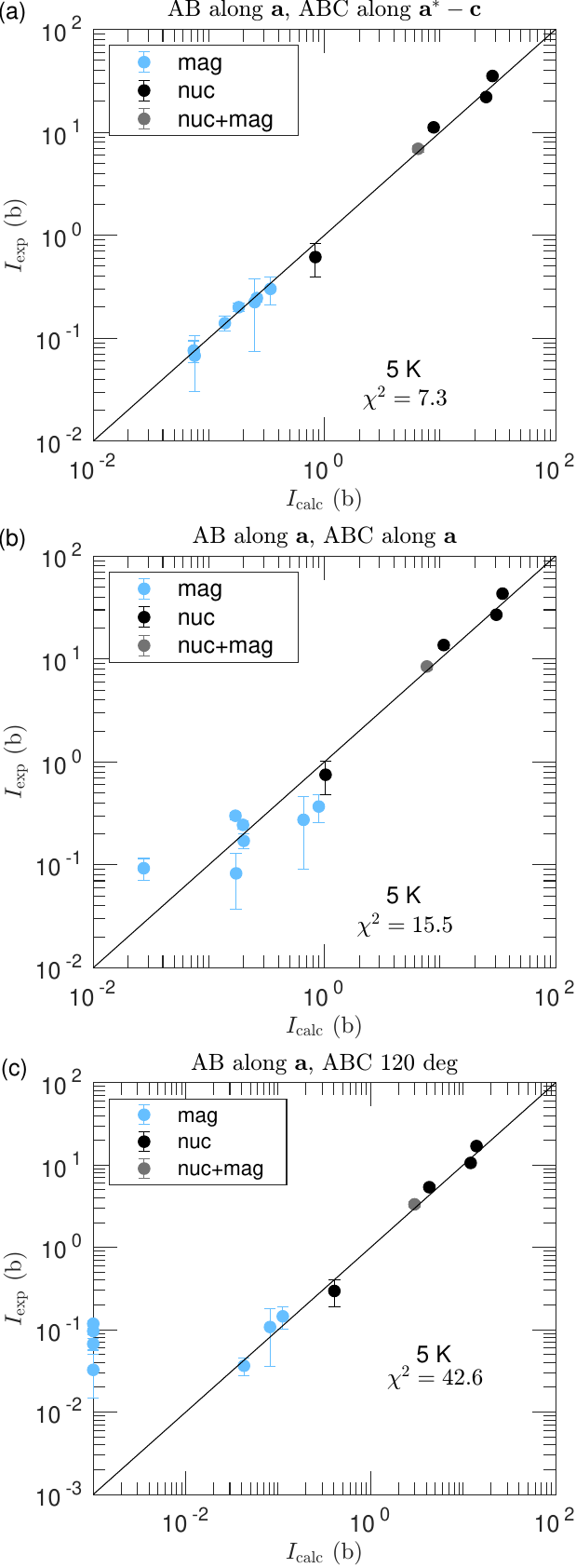}
    \caption{ Refinement of the single crystal diffraction data from the elastic line of SEQUOIA for \VNS. The figures show the refinements and chi-squared values for the assumption that $AB$ crystals have $a$-axis AFM order, while (a) assumes $ABC$ crystals have canted AFM order in the $\vb{a}^*-\vb{c}$ plane, (b) assumes $ABC$ crystals have pure $a$-axis AFM order, and (c) assumes $ABC$ crystals have $120^\circ$ order. Most magnetic peaks in the latter case would be theoretically absent, in contrast with experiment, so this structure cannot alone describe the data. }
    \label{fig:SI_Diffraction_Refinement}
\end{figure}

Neutron scattering probes the relative orientation of spins through the vector magnetic structure factor (Eq.~\ref{FmQ}). For single domain collinear magnetic orders the unique direction can be differentiated by looking at $(10l)$-type peaks. If the spins are arranged AFM between layers and oriented along the $c$-axis, then the magnetic diffraction intensity satisfies: 
 \begin{equation} 
     I_m(10l) \propto (1-\hat{Q}_z^2) m_z^2 |f(Q)|^2 
 \end{equation} 
where $\hat{Q}_z = Q_z/|\vb{Q}|$, and will therefore decrease in intensity for increasing $l$. %Indeed, the observed monotonically decreasing intensity of the $(10l)$ peaks for increasing $l$ indicates that the spins have a significant $\vb{c}$-axis AFM spin component. 
Contrastingly, had the spins been arranged AFM between layers and oriented along the $\vu{y} = \vu{a}^*$-axis, the intensity would instead behave as 
 \begin{equation} 
     I_m(10l) \propto \hat{Q}_z^2 m_y^2 |f(Q)|^2 
 \end{equation}
which for small $\hat{Q}_z, l$ would initially increase in intensity before decreasing due to the magnetic form factor $f(Q)$. The latter result changes slightly when considering domains. Throughout our analysis of diffraction and spectroscopy, we account for domain averaging. %Note if there are two orthogonal components of spin $\mu_y$ and $\mu_z$, then the structure factor is simply the sum of these contributions. 
Fig.~\ref{fig:SI_Diffraction_Refinement} shows the refinement of the multi-crystal ensemble (measured on SEQUOIA) for different models, with the best fit occurring for the spins along $\vb{a}^*-\vb{c}$. 

A set of two polytypes with $N_1 + N_2 = N$ number of unit cells amount to a superposition of the scattering intensity, $I = fI_1 + (1-f)I_2$, where $f=N_1/N$ is the molar fraction of polytype 1. Note that the unit cell volume for the $ABC$ polytype that we label as 2 is $v' \approx 3v$ as polytype 1 is $A\bar{B}$ with lattice parameter $c$, while polytype 2 has the stacking sequence $A\bar{B}C\bar{A}B\bar{C}$ with lattice parameter $c'=3c$. Considering the small number of magnetic peaks involved, we constrained the moment size for the two polytypes to be equal. We calculate the $\chi^2$ statistic and let all three parameters refine freely until they reach a threshold $\chi^2_{\mathrm{min}} (1 + 1/\nu)$ where $\nu=N_{\mathrm{data}}-N_{\mathrm{par}} = 9$. We find a molar fraction $f=0.45(7)$ for AB crystals in the ensemble. The in-plane and out-of-plane spin component of the $ABC$ crystal is fit as $m_y = 0.7(2)~\mu_B$ and $m_z = 1.3(1)~\mu_B$ respectively, amounting to the total moment size for both crystals as $m = \sqrt{m_y^2 + m_z^2} = 1.5(4)~\mu_B$. 

%From our experiment, we are not sensitive to potential relative lengths $\mu_1/\mu_2$. This is because the intensity is proportional to the product of the molar fraction and the moment squared, and thus changes in the molar fraction can be compensated by changing the spin ratio. 
%Chi-squared distribution for the refined magnetic moment $\mu$ per vanadium ion. The dashed line represents the rise of $\chi^2$ from varying parameters up to $\chi^2_{\mathrm{min}}(1+1/\nu)$ \cite{SM}. 

\subsection{Neutron spectroscopy} 
\begin{figure}
    \centering
    \includegraphics[width=0.8\linewidth]{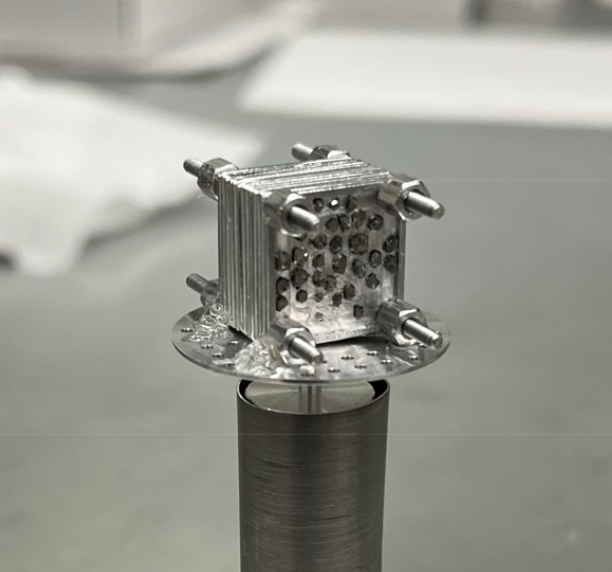}
    \caption{ Ensemble of co-aligned single crystals of \VNS, grown by the chemical vapor transport method, used for the experiment at SEQUOIA. }
    \label{fig:SI_Ensemble}
\end{figure}

Crystals of \VNS were grown by the chemical vapor transport as described elsewhere \cite{Ray2025}. The crystals tend to be no larger in dimension than 1 mm $\times$ 1 mm $\times$ 0.5 mm with an average mass less than 1 mg. While sufficient for neutron diffraction, their small volume has previously inhibited neutron spectroscopy, and to date there have been no reports of experimental investigations into the exchange interactions in \VNS. We co-aligned an ensemble of approximately 300 single crystals with a total mass of 328(1) mg, on an aluminum mount using minimal amounts of the fluoropolymer adhesive CYTOP (Fig~\ref{fig:SI_Ensemble}). Although we did not have a significant background from aluminum or CYTOP, we have an approximately constant background from the incoherent scattering of vanadium. With SEQUOIA, we measured the spectrum at 5~K using $20$~meV incident-energy neutrons. The multi-crystal sample was aligned in the $(hhl)$ scattering plane with $(1\bar{1}0)$ perpendicular to the scattering plane. Our ensemble had a mosaic of 2.4(1)$^\circ$ as determined by rocking scans on the $(002)$ nuclear Bragg peak about the $(1\bar{1}0)$ axis. We symmetrize with the operations of space group $P6_{3}22$ to optimize statistics. The amplitude in the refinement of the nuclear diffraction data was used to convert count rates to an experimental measure of the inelastic magnetic neutron scattering cross section per formula unit.  

We fit the exchange constants to the data using linear spin-wave theory with the SpinW package \cite{toth_linear_2015}. We partitioned the data into voxels of equal linear dimensions $\delta Q = (2\pi/c)0.12 \approx 0.06$ \AA$^{-1}$ and energy bins $\delta (\hslash\omega) = 0.1$ meV. Given a set of Hamiltonian parameters, we self-consistently deduce the ground state by optimizing the ground state given an initial guess of the proposed $A$-type magnetic structure. The excitation spectrum was then computed given this spin structure. The model parameters were optimized to achieve the closest agreement with the data set acquired for the magnetic neutron scattering cross section. To avoid the contribution from the elastic line, we excluded data with $\hslash\omega_j\le 1.3$ meV. Because of the small sample and the large number of pixels, most pixels have a small event count that is insufficient for the Poisson distribution to be approximated by a normal distribution. Instead of least-squares minimization, we therefore minimize the reduced negative log-likelihood \cite{Reeder2025} 
 \begin{equation} 
    L(\vb{x}_i, \vb*{\theta}) = - \frac{1}{N} \sum_i^N \ln \frac{\lambda_i^{k_i} e^{-\lambda_i} }{k_i!} 
 \end{equation} 
This corresponds to identifying the model parameters that maximize the probability of obtaining the observed event count in each of the $N=1,973,283$ pixels. Here $\vb{x}_i = (\vb{Q}_i, \omega_i)$ is a point in momentum-energy space, $k_i = n_i I(\vb{x}_i)$ is the detected number of counts, $\lambda_i = n_i I_\mathrm{th}(\vb{x}_i)$ is the number of counts predicted by the model, and $n_i$ is the instrument normalization. The intensity is calculated as $I_{\mathrm{th}}(\vb{x}_i) = I_{\mathrm{LWST}}(\vb{x}_i)/A + I_0$ where $I_{\mathrm{LSWT}}(\vb{x}_i)$ is the LSWT intensity in absolute units, $A\equiv1$ is the normalization (given the absolute normalization to the nuclear diffraction data), and $I_0$ is the constant background from the incoherent scattering from vanadium. The likelihood statistic originates from the Poisson distribution of detected neutron counts. The confidence region is the range of parameters such that $L$ increases by $1/(2\nu)$ where $\nu=N-N_\mathrm{param}$ is the number of degrees of freedom \cite{Frodesen1979}. Since $\nu$ is exceedingly large, $1/(2\nu)$ becomes vanishingly small, whereas the uncertainty on the parameters may be larger.

To get a better representative error estimator based on the limitations of our data, we calculate $L_p$ when taking Poisson-random numbers $\lambda_i = \mathrm{Poisson}(\lambda_i)$, forming a distribution $P(L_p)$ of $L_p$ for hundreds of Poisson random values. The resulting distribution is approximately normal with FWHM approximately $\sigma_L \approx 1/900$. Hence, we approximate the error bars on the parameters as $1\sigma$ (standard deviations) for which $L$ rises by $\sigma_L$. The parameters are varied by a Monte Carlo method using principal component analysis, as described elsewhere \cite{Scheie2022, Lygouras2024}. 

\begin{figure}
    \centering
    \includegraphics[width=\linewidth]{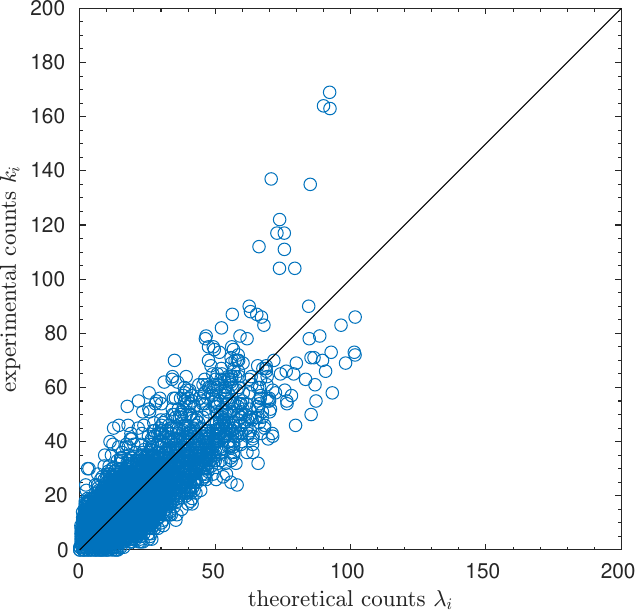}
    \caption{ The result of the refinement of the inelastic spectrum for \VNS, which shows the theoretical Poisson count $\lambda_i$ versus the experimental count $k_i$ for pixel $i$. Using the negative log-likelihood statistic, the minimization routine attempts to optimize the count rate with the theoretical expected value $k_i = \lambda_i$. } 
    \label{fig:PoissonCountFit}
\end{figure}

The fit result for the theoretical counts $\lambda_i$ versus the experimental counts $k_i$ are shown in Fig.~\ref{fig:PoissonCountFit}. The counts indeed fall on the line close to the expected line $k_i = \lambda_i$, demonstrating the goodness of fit. 

\subsection{Spin wave analysis} 
The $A$-type AFM in the $AB$ polytype has a magnon spectrum that is established in the literature within linear spin wave theory \cite{Huberman2005, Boothroyd2020}. We define the $x$-axis to be the quantization axis for which $S^x_A = S - a^\dagger a$ and $S^x_B = -S + b^\dagger b$. We define $\hat{S}_A^\pm = \sqrt{2S} a^\pm$ and $\hat{S}_B^\pm = \sqrt{2S} b^{\mp}$ for which the Heisenberg Hamiltonian is written as a quadratic form with the following dynamic matrix: 
 \begin{align} 
   \mathsf{H}_{\vb{q}} &= \mqty( A_{\vb{q}} & 0 & C_{\vb{q}} & D_{\vb{q}} \\ B^*_{\vb{q}} & A_{\vb{q}} & D^*_{\vb{q}} & C_{\vb{q}} \\ C_{\vb{q}} & D_{\vb{q}} & A_{\vb{q}} &0 \\ D^*_{\vb{q}} & C_{\vb{q}} & 0 & A_{\vb{q}} ) \\ 
   A_{\vb{q}} &= [J_A(\vb{q}) - J_A(0)] - |D| + J_{AB}(0) \\
   C_{\vb{q}} &= |D| \\
   D_{\vb{q}} &= J_{BA}(\vb{q}) = J_{AB}^*(\vb{q}) 
 \end{align}
Here $J_A(\vb{q})$ and $J_{AB}(\vb{q})$ describe the Fourier transforms of interactions on the same sublattice or different sublattice, respectively, where 
 \begin{equation}
     J_{dd'}(\vb{q}) = \sum_{j} J_{idjd'} e^{i\vb{q}\cdot \vb{R}_{idjd'}} 
 \end{equation}
Diagonalizing the Hamiltonian matrix $\mathsf{H}_{\vb{q}}$ while respecting the symplectic commutation relation $[\vb{X}_{\vb{q}}, \vb{X}^\dagger_{\vb{q}'}] = \mathrm{diag}(I,-I) \delta(\vb{q}-\vb{q}')$ requires the Bogoliobuv transformation $\vb{Y}_{\vb{q}} = P_{\vb{q}} \vb{X}_{\vb{q}}$ which yields the (positive) energy eigenvalues of the spin waves, \cite{Boothroyd2020} 
 \begin{equation}
     \hslash\omega_\pm(\vb{q}) = S\sqrt{A_{\vb{q}}^2 - C_{\vb{q}}^2 \pm 2C_{\vb{q}} \sqrt{|D_{\vb{q}}|^2} - |D_{\vb{q}}|^2}
 \end{equation}
%The dispersion in these modes will generically have one band with a Goldstone mode (corresponding to the state where all spins are rotating in the basal plane) and one gapped mode (where the spins are rotating out of the basal plane at the cost of the single-ion energy $D_z$). 

\begin{figure}
    \centering
    \includegraphics[width=\linewidth]{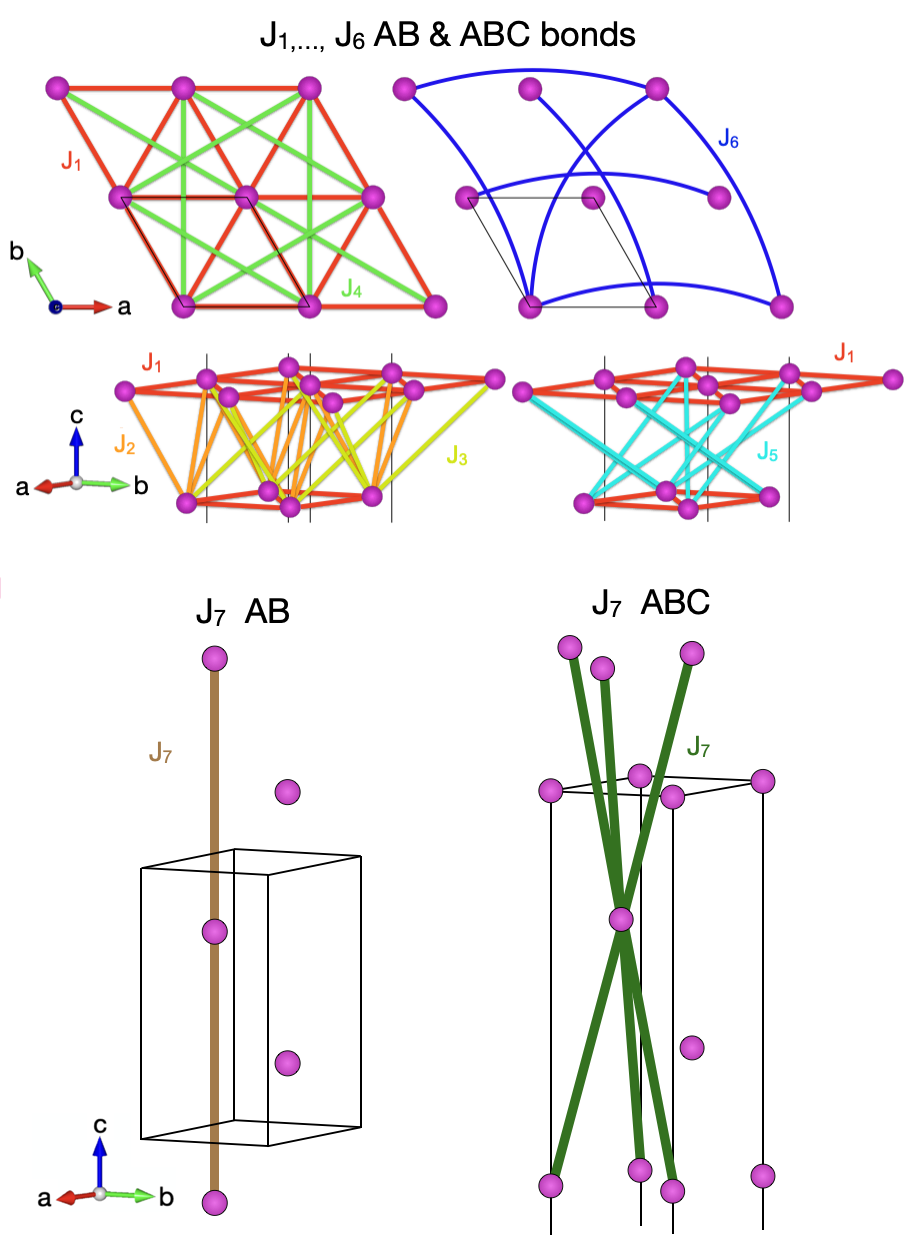}
    \caption{ Visualization of the bonds $J_1$,...,$J_7$ in \VNS for the $AB$ and $ABC$ polytypes considered in this work. The V atoms are shown in purple. For the first six bonds, the coordination numbers are the same among the two polytypes. However, the largest bond, $J_7$, is the first bond to differ between the two polytypes, due to the different stacking. This, obviously, results in differing geometry.}
    \label{fig:ExchangeBonds}
\end{figure}

The respective bonds considered in this work are shown in Fig.~\ref{fig:ExchangeBonds}.
%and Fig.\ref{fig:ExchangeBondsJ7}. 
It is worth highlighting the bandwidth at relevant high-symmetry points such as $A$ and $K$ respectively: 
 \begin{gather}
    \hslash\omega_A \equiv \hslash\omega(00\tfrac{1}{2}) = 2S(3J_2 + 3J_3 + 6J_5 - 2J_7) \approx 5~\mathrm{meV}\\ 
    \hslash\omega_K \equiv \hslash\omega(\tfrac{1}{3} \tfrac{1}{3} 0) = 3S (-3J_1 + 2J_2 + 2J_3 + 4J_5 - 3J_6) \approx 8~\mathrm{meV}
 \end{gather} 
when we treat $D$ as negligible. These highlight the approximate maximal bandwidths of the magnons observed in the experiment, and detail how the different interaction strengths compete. We note that only inter-planar interactions such as $J_2$ and $J_3$, etc. contribute to the $A$-point magnon energy. 

%\begin{figure}
%    \centering
%    \includegraphics[width=\linewidth]%{Figures_DFT/ExchangeBonds_improved.pdf}
%    \caption{ Visualization of the bonds $J_1,\dots J_6$ in \VNS considered in this work. The \ce{V} atoms are shown in purple. For these bonds, the coordination numbers are the same among the two polytypes.}
%    \label{fig:ExchangeBonds}
%\end{figure}

\subsection{RKKY interactions}
The prefactor $A=310(50)$~$\mathrm{meV}\cdot$\AA$^4$, estimated from the fit to the RKKY function $J(R) = Ak_F^4 F(2k_F R)$, is used to estimate the Kondo coupling through the relation \cite{Ruderman1954}
\begin{equation} \label{eq:RKKY_Amplitude} 
    Ak_F^4 = \frac{J_K^2 v_0^2 k_F^4 m^*}{4(2\pi)^3 \hslash^2} = \frac{(3\pi^2)^2}{8(2\pi)^3} \frac{J_K^2}{(\hslash^2 k_F^2/2m^*)} = \frac{9\pi}{8^2} \frac{J_K^2}{E_F} 
\end{equation} 
where $J_K$ is the Kondo coupling strength, $v_0=V/N$ is the atomic volume, $k_F^3/3\pi^2 = N/V$ under the free-electron model, and $E_F$ is the Fermi energy. With $E_F=\hslash^2 k_F^2/2m^*$, and $m^*$ equal to the bare-electron mass, and $k_F$ extracted from the RKKY function fit for self-consistency, we find $J_K$~$= 91(7)$~meV. 

We have also considered an anisotropic RKKY model for an ellipsoidal (more specifically, an oblate spheroidal) Fermi surface. This arises due to the effective mass anisotropy $\alpha=m_\parallel/m_\perp$, where $m_\perp (m_\parallel)$ represents the effective mass for electron motion along $k_x(k_z)$. Here the Fermi surface is given by the equation $E_F = \frac{\hslash^2}{2} \left( \frac{k_\perp^2}{m_\perp} + \frac{k_\parallel^2}{m_\parallel} \right)$, and in the free-electron model, $k_\parallel/k_\perp = \alpha^{1/2}$ for the semi-axes of the ellipse (the respective Fermi momenta). The derivation \cite{Story1992} for the exchange coupling in this case is nearly identical for that of the spherical Fermi sea, with the argument $2 k_F R$ replaced by $2k_\perp R_\alpha$. Here $k_\perp$ is the in-plane Fermi momentum and $R_\alpha = \sqrt{R_x^2+R_y^2+\alpha R_z^2}$. Thus, the main effect of the effective mass anisotropy on the oscillations in the exchange constants is to change the effective distance of the bonds, due to the distortion of the Fermi sea. In the limit of $\alpha=m_\parallel/m_\perp \to \infty$, the oscillations decay rapidly for out-of-plane bonds, leaving an oscillatory character for in-plane bonds. This limit is to be expected for systems with quasi-two-dimensional Fermi surfaces, as in the cuprates. In the limit $\alpha\to 0$, the Fermi surface has flatter regions along $k_z$, leading to enhanced oscillations for out-of-plane oriented bonds. 

If taking $\alpha\ll1$ as a small value to match the approximately-ellipsoidal Fermi surface near $\Gamma$ with the small aspect ratio, the best fit value comes out to $k_{F\perp}=0.83$\AA$^{-1}$ and $\alpha=0.05$. Note that $k_{F\perp}=0.83$\AA$^{-1}$, closest in value to $|K|=|K'|=4\pi/3a \approx 0.73$ \AA$^{-1}$, is greater in magnitude than the extent of the Brillouin zone. That $2k_{F\perp}\approx |K-K'|$ in this assumption is suggestive of intraband scattering between approximately-cylindrical Fermi surfaces. However, such cylindrical surfaces would have large $\alpha\gg 1$, in contradiction to the assumption that $\alpha\ll 1$. 

Conversely, if taking $\alpha \gg 1$ for scattering from highly cylindrical Fermi surfaces, the fit suffers due to the predicted rapid decay of the out-of-plane interactions due to the large value of $R_\alpha$. In contrast, the largest interactions we find in our experiment are for out-of-plane interactions. 

\begin{figure}
    \centering
    \includegraphics[width=\linewidth]{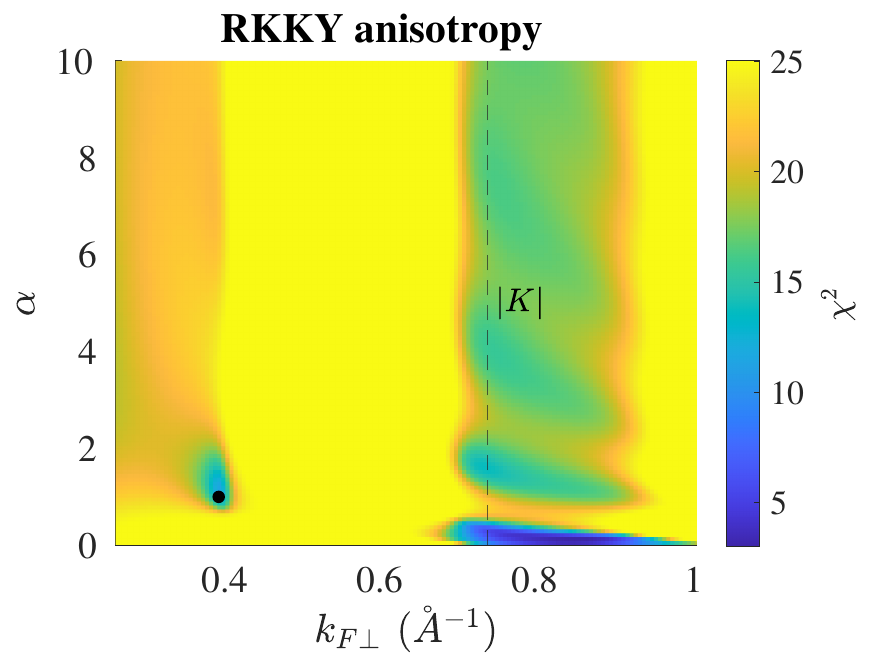}
    \caption{ Chi-squared minimization of the anisotropic RKKY model for \VNS, with varying anisotropy factor $\alpha$ and in-plane Fermi wavevector $k_{F\perp}$. The largest extent of the first Brillouin zone is $|K|=4\pi/3a$. }
    \label{fig:RKKY_anisotropy} 
\end{figure}

For general $\alpha$ and $k_{F\perp}$, we can plot the chi-squared metric (having optimized $A$ in Eq.~\ref{eq:RKKY_Amplitude} at each point). We find that the best fit that meets the above constraints holds near $\alpha=1$. That is, for our fits to the general anisotropic Fermi surface model, there is no self-consistent solution to relate to the calculated Fermi surface sheets. In this regard, we treat the isotropic RKKY model as a natural approximation of the complicated Fermi surface with multiple sheets, simultaneously allowing us to gauge the effective size of the Fermi surface while also giving access to the strength of the Kondo coupling between localized and itinerant spins.

\subsection{Luttinger-Tisza theory} 
\begin{figure}
    \centering
    \includegraphics[width=\linewidth]{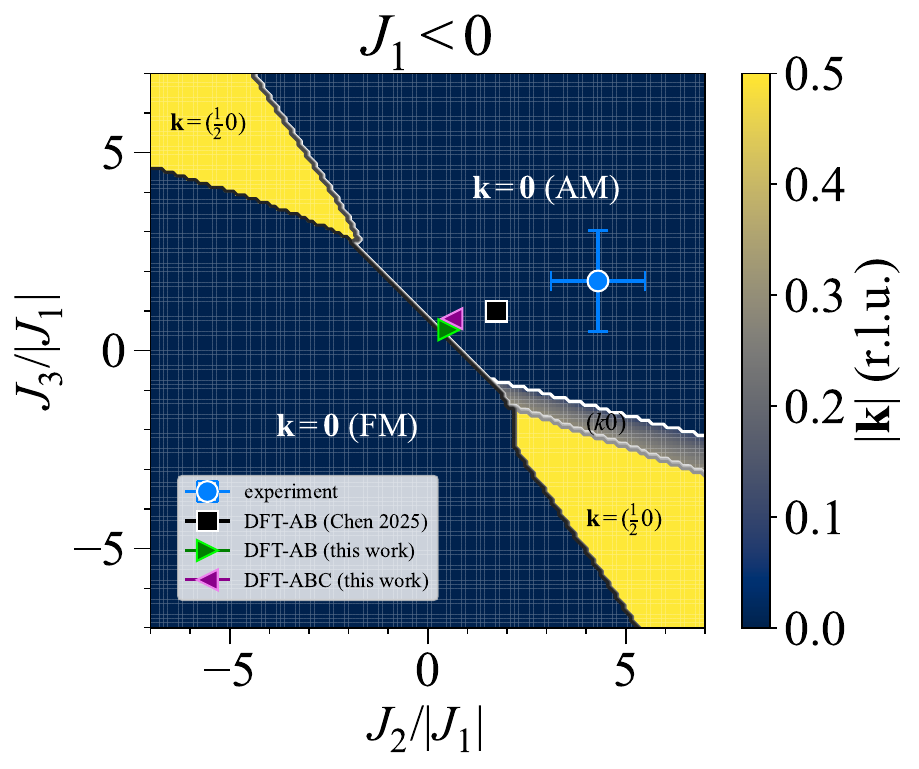}
    \caption{ Luttinger-Tisza model for \VNS. We determine the predicted ground-state spin structures and wavevectors (colored regions) for varying exchange constants $J_2/|J_1|$ and $J_3/|J_1|$, with other values fixed from experiment. Additionally shown are the values from DFT calculations. It can be seen that all values are consistent with the altermagnetic order. Furthermore, the $\vb{k}=\vb{0}$ altermagnetic (AM) magnetic ground state is stable over a wide range of parameter variation.}
    \label{fig:SI_LuttingerTisza}
\end{figure}

Having calculated the relevant exchange tensors for \VNS, we predict the ground state using the methods of Luttinger-Tisza \cite{friedman_solution_1974}. The isotropic Heisenberg interactions with in-plane easy axis anisotropy means we can simply evaluate the eigendecomposition of the exchange tensor 
 \begin{equation}
     \mathcal{J}(\vb{q}) = \mqty( J_A(\vb{q}) & J_{AB}^*(\vb{q}) \\ J_{AB}(\vb{q}) & J_A(\vb{q}) ) 
 \end{equation} 
The eigenvalues are $\lambda_{\vb{q}}^{\pm} = J_A(\vb{q}) \pm |J_{AB}(\vb{q})|$ and the eigenvectors are $\vb{v}^{\pm}_{\vb{q}} = (\pm\sqrt{J_{AB}^*(\vb{q})/J_{AB}(\vb{q})}, 1)$. The eigenvector gives information of the relative spin orientations. The minimum energy solution occurs for the negative signs. For a simple $J_1-J_2$ model, the solution is $\vb{v}^{-}_{\vb{q}=\vb{0}} = (-\mathrm{sgn}(J_2),1)$ which corresponds to AFM (FM) alignment on the two sites if $J_2>0$ ($J_2<0$). In general, we can consider the best solution from our refinement of INS data and vary parameters like $J_2/|J_1|$ and $J_3/|J_1|$. Fig.~\ref{fig:SI_LuttingerTisza} shows the ground-state landscape, and notably the $\vb{k}=\vb{0}$ AFM order is stabilized over a wide range of exchange parameters. 

Importantly, while there is a DM vector allowed by symmetry between all ions in the unit cell, the Fourier transform
 \begin{equation}
   \vb{D}_{\vb{q}} = \sum_j \vb{D}_{id,jd'} e^{i\vb{q}\cdot \vb{R}_{idjd'}}
 \end{equation}
vectorially sums to zero when $\vb{q}=\vb{0}$, implying that the interaction tensor $[\mathcal{D}(\vb{q})]_{\alpha\beta} = \epsilon_{\sigma\alpha\beta} D_{\vb{q}}^\sigma$ is likewise identically zero. Hence, for the A-type AFM order, the DM vector can not be responsible for the spin canting out-of-plane. 

\subsection{Magnetization and mean-field theory} 
We use a mean-field treatment for the $S=1$ spins to demonstrate the agreement of our exchange constants with the thermodynamics of \VNS. We deduce the Curie-Weiss temperature from the exchange constants \cite{Lee2014} as $k_B \Theta_{\mathrm{CW}}$~=~$-\tfrac{S(S+1)}{3} \sum_k z_k J_k$ where $z_k$ is the coordination number for the $k$th near-neighbor. To estimate the transition temperature, we write the mean-field Hamiltonian $\mathcal{H}_{\mathrm{mf}}$~=~$\sum_i \bar{J}\expval{S_i^x}$ where $\bar{J}$~=~$\sum_k z_k \sigma_k J_k = (6J_1-6J_2-6J_3 +\cdots) < 0$ and $\sigma_k$~=~$\expval{S_k}/\expval{S_1}$ is the relative sign of neighboring spins (listed in Table~\ref{tab:J_couplings_DFT}). Given a mean-field Hamiltonian $\mathcal{H}_i = h_i^x \hat{S}_i^x$ with mean-field $h_i^x = \sum_j J_{ij} \expval{\hat{S}_j^x}$ (assuming negligible anisotropy), the mean-field transition occurs when $k_B T_N = -\frac{S(S+1)}{3} \bar{J}$. Quantum fluctuations for small $1/S$ will generally result in a smaller $T_N$ than the mean-field value. However, this triangular lattice magnet is not expected to have frustrated interactions given the small frustration index $|\Theta_{\mathrm{CW}}|/T_N \sim 1$. Within this mean-field model, we predict a Curie-Weiss temperature $\bar{\Theta}_{\mathrm{CW}}$~=~$-50(20)$~K and Neel temperature $\bar{T}_N$~=~$61(2)$~K. These are in the same order of magnitude as the experimental values of $\Theta_{\mathrm{CW}}$~=~$-70$~K and $T_N$~=~$50$~K respectively \cite{Ray2025}. 

\begin{figure}
    \centering
    \includegraphics[width=\linewidth]{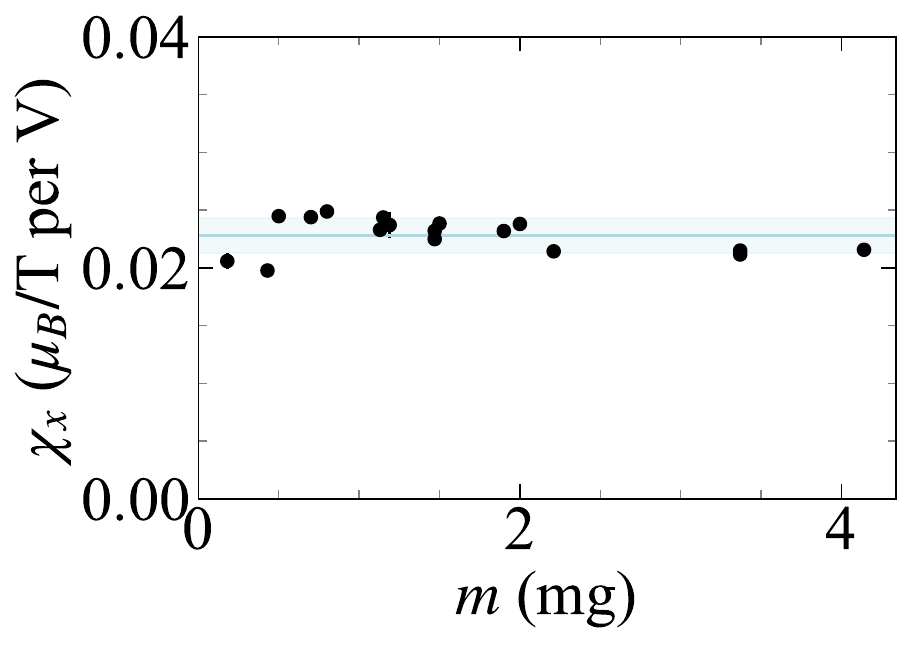}
    \caption{In-plane magnetic susceptibility $\chi_x = \dv*{M_x}{H}$ for \VNS in numerous samples with different masses. Such samples were measured as-grown or after mechanical polishing. Despite sample variation (from stoichiometry or growth conditions), there is a uniform distribution with average $\bar{\chi_{x}} = 0.023(1)\mu_B/$~T. }
    \label{fig:VNS_susceptibility} 
\end{figure}

In the simplest mean-field model for $S=1$ spins, we study the effect of a magnetic field applied perpendicular to the ordered moments. For small fields in a single-domain magnet, the magnetic field will slightly polarize the spins along the applied field direction, resulting in an in-plane magnetic susceptibility 
\begin{equation} 
    \chi_x = \dv{M_x}{H}\bigg|_{H\to 0} \approx \frac{\sqrt{3}}{2} \frac{(g\mu_B)^2}{(-\bar{J})} \approx 0.026~\mu_B/\mathrm{T}. 
 \end{equation} 
The limit holds when $k_B T\ll |\bar{J}|$ where $(-\bar{J}) = \tfrac{3}{2} k_B T_N$ in the mean-field model. This result is remarkably similar to the experimental susceptibility, with an average of $\bar{\chi}_x$~=~$0.023(1)~\mu_B/\mathrm{T}$ (see Fig.~\ref{fig:VNS_susceptibility}). In our mean-field approach, we have assumed that $\bar{J}\sim T_N$ is approximately equal for all samples. Furthermore, in this mean-field approximation, full saturation is reached when the applied magnetic field strength $H^*$ is enough to overcome the exchange field, $h_i = g\mu_B S H^*$. This is predicted to occur at a critical field $H^* = (-\bar{J})/(\sqrt{3} g\mu_B/2)\approx 80$~T.

We found that some samples have a finite remnant magnetization in addition to the linear slope. Consequently, we fit the magnetization data in Fig.~\ref{fig:UniversalHall}(b) with a sum of two components: A linear term associated with magnetizing the bulk AFM plus a phenomenological non-linear term to extract the remnant magnetization 
 \begin{equation} \label{eq:TwoFunctionFits}
     \tilde{M}_x(H) = a H + b\left( \tfrac{2}{\pi}\arctan(k[H-H_c]) \right) 
 \end{equation} 
Here $a$ represents the susceptibility, $b$ represents the magnitude of the remnant magnetization, and $H_c$ represents the coercive field. Other functions, such as $\tanh$, are possible, but the curvature near the low-field regime appears to match best with Eqn.~\ref{eq:TwoFunctionFits}. Based on the assumption that $\chi_x$ is sample independent, the following estimate for in-plane remnant magnetization: $\mu_0 \Delta M_x = \mu_0 \rho \bar{\chi}_x b/a$, where $\rho$ is the theoretical number density of vanadium per unit cell was used to eliminate the considerable error bar associated with mass measurements on the very small samples employed. 

\begin{figure}
    \centering
    \includegraphics[width=\linewidth]{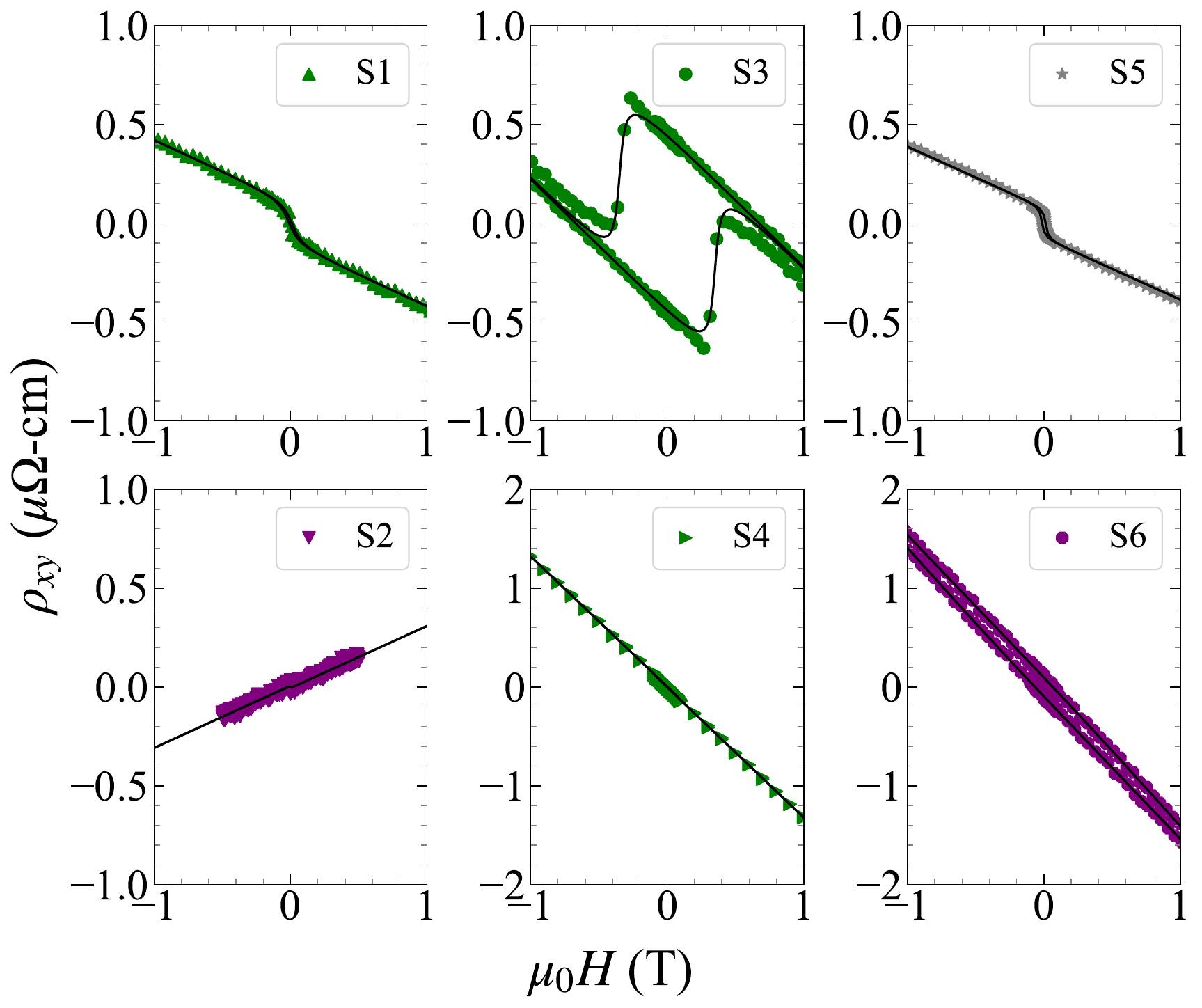}
    \caption{ Hall effect signals for various samples S1-S6 of \VNS. The top (bottom) row shows samples with significant (marginal) AHE $\Delta \rho_{xy}$. When explicitly measured with diffraction, green (purple) refers to samples with predominantly $AB$ ($ABC$) stacking. Note the change in sign of both the Hall coefficient $R_H$ and the AHE $\Delta \rho_{xy}$ among samples. Black lines are fits to Eq.\ref{eq:TwoFunctionFits}. The charge carriers are electrons for the majority of measured samples. }
    \label{fig:SI_Hall} 
\end{figure}

A similar fitting procedure is done for $\rho_{xy}$ to derive $R_H$ and $\Delta \rho_{xy}(H\to0)$, where $R_H = d\rho_{xy}/dB$ is the Hall coefficient. In Fig.~\ref{fig:SI_Hall}, we show the Hall resistivity of various samples S1-S6 used to extract $\Delta \rho_{xy}$ in Fig.~\ref{fig:UniversalHall}. The fits to the functional form of Eq.\ref{eq:TwoFunctionFits} are shown with black lines. With $R_H=1/ne$ in a one-band model, $R_H$ can in principle vary among samples due to varying chemical potential. However, it is worth noting that the average carrier density among the measured samples (in this work and in the literature), $\bar{n}~=~1.1(8)\times 10^{27}$~m$^{-3}$, is comparable to that expected from the free-electron theory using $k_F$ from the RKKY fit: $n = k_F^3/3\pi^2 = 1.85\times 10^{27} $~m$^{-3}$. 

%The discrepancy between the two polytypes in our model occurs because by $n=7$ near-neighbor, the coupling changes between AA sites in the $AB$ crystal $(z_7=2)$, to AC sites in the $ABC$ crystal $(z_7=6)$. In this approximation, the discrepancy in transition temperature is $\delta T_N \approx -2(\delta z_7) J_7/3k_B$. 
%In our model, there is a discrepancy in $T_N$ due to the distinct bonds and coordination numbers in the two polytypes (see Fig.~\ref{fig:Diffraction}(d)). This amounts to a difference $\delta T_N \approx 8(1)$~K, between the ABC and AB polytypes, which explains the magnitude of the discrepancy but not the sign. On the basis of our DFT calculations, the likely origin of this discrepancy comes from the differences in the interaction strengths in the two polytypes. 

%\subsection{Raman scattering} 

\subsection{Statistical defects} 
Anomalous Hall resistivity driven from Weyl nodes is an intrinsic property of the sample due to enhanced Berry curvature. However, extrinsic factors, such as disorder or domains, can reduce this intrinsic quantity as a result of reduced phase coherence between layers. To see this, consider a crystal of thickness $t$ made up of $N$ atomically thin slabs with thickness $\delta t$. Assuming that each layer of the crystal has a fixed lateral area and resistivity, then the input current $I_x$ is evenly divided between the $N$ layers (each carrying current $I_x/N$, as in a parallel circuit). The Hall voltage in metals $\Delta V_y$ depends on the average macroscopic electric field $\vb{E}_y$ generated in the sample \cite{Smejkal2022}, and thus is expressed as an average over the voltage $\Delta V_y^{(k)}$ in each layer $k$: $\Delta V_y = (1/N) \sum_{k} \Delta V_y^{(k)} = I_x (\rho_{xy}/t)$. However, microscopically, we can consider the electric fields being generated in each layer of thickness $\delta t$. If each layer has a Hall resistivity $\rho_{xy}^{(k)}$, the Hall voltage in each layer is $\Delta V_y^{(k)} = (I_x/N) \cdot [\rho_{xy}^{(k)}/(\delta t)] = I_x\rho_{xy}^{(k)}/t$. This implies that the Hall resistance observed is the average among layers: $\rho_{xy} = (1/N) \sum_k \rho_{xy}^{(k)}$. As an example, in a normal metal under a magnetic field $B$, the Hall resistivity is $\rho_{xy} = B/nq$ where $n$ is the density of carriers with charge $q$. Since there is phase coherence between the layers, $\rho_{xy}^{(k)} = \rho_{xy} = B/nq$. 

When $\rho_{xy}$ is driven by Berry curvature instead of an external magnetic field, there can be changes (in sign or magnitude) to the Hall resistivity between layers. To model the anomalous Hall resistivity in \VNS, we consider $N$ layers of $A$, $B$ or $C$ sites, each layer having an intrinsic anomalous Hall resistivity $\rho_{xy}^{(k)} \propto h_k$, where $h_k$ represents the intrinsic contribution proportional to the Hall vector. We suppose that $h_k$ may vary among layers in sign, based on its connection to the magnetic structure. In what follows, $h_k$ is normalized such that, if the Hall vectors are coherent among all layers, then $\expval{h}\to 1$. We calculate the sample-averaged value $\Delta \rho_{xy} \propto \expval{h} = (1/N) \sum_k h_k$. Similarly, to model the in-plane magnetization, we attribute each layer as having a net magnetic moment $m_x^{(k)}$ (proportional to the net spin in each layer) and calculate the average $\expval{m_x} = (1/N) \sum_k m_x^{(k)}$. Namely, $\expval{m_x}\to 0$ in a perfect AFM.
%As a first approximation we assume that the resistivity $\rho_{xx}$ is approximately constant between samples, allowing for an identification of $\expval{h} \propto \rho_{xy}/\rho_{xx}^2$ in order to compare with experiment. 

We first consider a model of domains with disparate signs of the Hall vector driven from stacking faults. In an AB-stacked crystal, domains with a within a single layer, vanadium sites can occupy either $A$ or $B$ positions. A misplaced site will result in a stacking pattern like $\cdots A\bar{B}A\bar{B}|B\bar{A}B\bar{A}\cdots$. If, in such $A\bar{B}-B\bar{A}$ structural domain walls, the occluding $\bar{B}-B$ sites interact ferromagnetically, then the in-plane spins switch signs between layers: one such state is $\ket{\cdots \rightarrow \leftarrow \rightarrow | \rightarrow \leftarrow \rightarrow \cdots}$. There is now a net in-plane FM moment if the adjoining domains have an odd number of spins. Additionally, this domain wall acts as a product of a mirror operator and a time-reversal operator $m_z \Theta$; this retains the sign of the in-plane magnetization but switches the sign of the Hall vector (along with the canted $c$-axis component of spin) across the domain. This can also be appreciated by noting that the bar switches from $A$ to $B$ planes across the structural domain wall. In this state, the set of Hall vectors is $\ket{\cdots \uparrow \uparrow \uparrow | \downarrow \downarrow \downarrow \cdots}$. We simulate the effect of such structural defects on the Hall resistivity $h$ and in-plane magnetization. We assume a set of $N\gg 1$ layers, which a good assumption for our samples since the typical crystal thickness is $t\sim 100$ $\mu$m $\sim 10^4 c$. If there are $1 < n < N$ domain walls in the crystal, we find heuristically that the net in-plane magnetization $\expval{m_x} \sim \sqrt{n}/N$; this is expected for a random-walk process where, at each domain wall, the magnetization changes by $\pm 1$. Meanwhile, we find that the average Hall vector goes as $\expval{h} \sim 1/\sqrt{n}$. That is, more domain walls will tend to decrease the average anomalous Hall resistivity due to partial cancellation of the Hall signal by the generation of more sign changes. If the expected number of defects $n$ in a crystal of $N$ layers follows a binomial distribution with probability $p$ of having a defect on any given layer (which may be set, for example, by the crystallization temperature), then by the central limit theorem $n \sim pN$. In this case, we have that $\expval{h} \sim \frac{1-p}{p} \expval{m_x}$, suggesting stacking disorder enhances both in-plane magnetization and Hall resistivity. 

The above case is an example of disordered stacking within $AB$ crystals. Another possibility is a domain formation of $AB$ and $ABC$ stacking, as observed in the work of Fender \textit{et al}. \cite{Fender2025}. We can likewise model this scenario by partitioning the crystal into segments of $AB$ or $ABC$ slabs. We assume that the easy axis anisotropy in the slabs changes akin to the bulk counterparts. This model assumes that the effective correlation length for easy-axis or easy-plane anisotropy is sufficiently large that the $AB$ and $ABC$ domains retain their bulk easy axes. We consider a model of domain formation where the expected density of $ABC$ slabs is $p$. With the introduction of $ABC$ domains within $AB$ crystals, there is a freedom for the in-plane spins to change their Neel vector between the $AB-ABC-AB$ domain boundary; it is indeterminate. This is because, as discussed previously, the $ABC$ grains produce $\rho_{yz}$ but not $\rho_{xy}$ AHE. Furthermore, the interaction energy $J_2\vb{S}_i\cdot\vb{S}_j \approx 0$ for spins across a domain wall given the disparate easy axes. Thus, there can be in general a lack of phase coherence of the Hall vector between $AB$ domains sandwiching an $ABC$ domain, in the sense that the Hall vector can randomly change signs across the $ABC$ domain. In our model, we randomly select the sign of the Hall vector across $AB$ domains, and once this is chosen, the sign of the Neel vector is known. From this, we can compute the average Hall resistivity and magnetization for various configurations. In our simulation, we find that, independent of $p$, the in-plane magnetization in this model follows $\expval{m_x} \propto 1/N$. In addition, for general $p$, we find that the Hall vector follows $\expval{h} \approx (1-p) + \alpha(p-p_c) N^{-1/2}$ where $p_c=\tfrac{1}{2}$. For example, as $p\to 0$ there are only $AB$ grains, and $\expval{h}\to 1$ in the limit of $N\to \infty$, as we expect. Taken together, we have $\expval{h} \sim (p-p_c) \expval{m_x}^{1/2}$. When $p$ is small, the anomalous Hall signal thus decreases with increasing magnetization. %This general trend is what we find experimentally, although the exact functional dependence is difficult to ascertain reliably due to the variance of the number of layers in each crystal, in addition to the distribution of possible fractions of grains $p$. 

\begin{figure}
    \centering
    \includegraphics[width=\linewidth]{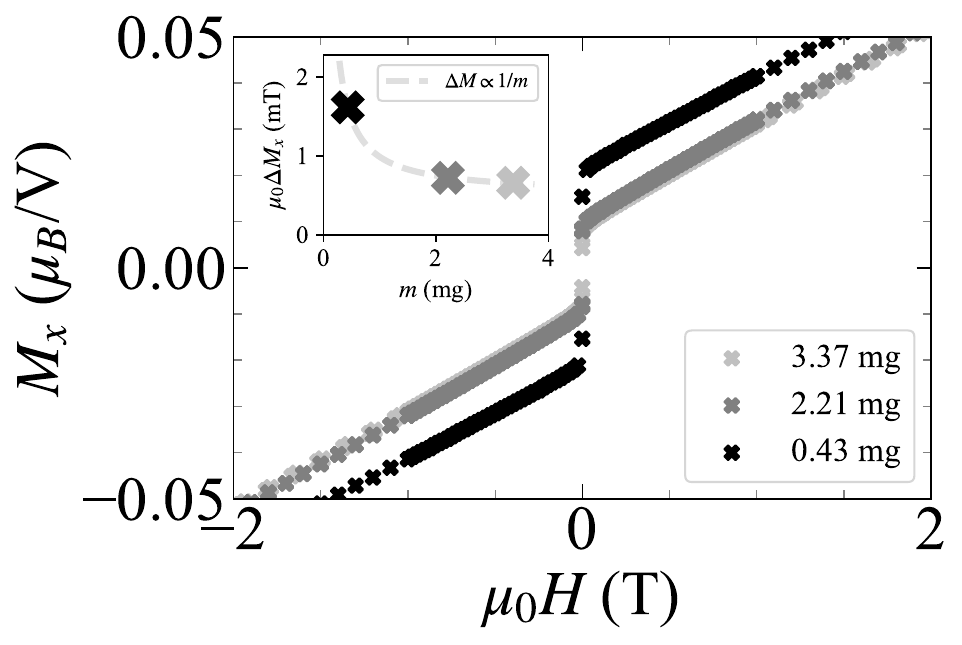}
    \caption{Effect of mechanical polishing in sample S7 (shown in Fig.~\ref{fig:UniversalHall}) of \VNS. Magnetization is an intensive quantity, yet polishing can affect the remnant magnetization in some samples. We can account for this in our model of AB-ABC domains. Error bars are smaller than marker sizes.} 
    \label{fig:VNS_polishing}
\end{figure} 

We use these models to obtain a qualitative sense of the changes of Hall resistivity with in-plane magnetization in Fig.~\ref{fig:UniversalHall}. Next, to get a sense of the accuracy of these predictions, we measure the magnetization of several samples before and after mechanical polishing. Some samples show virtually no change to the in-plane magnetization, which is consistent with $ABC$ samples having fixed in-plane spin canting. We find some samples with enhanced in-plane remnant magnetization after mechanical polishing (Fig.~\ref{fig:VNS_polishing}). The fact that $\Delta M_x\propto 1/m$, where $m$ is the sample mass, is in line with the prediction that $\Delta M_x \propto 1/N$ assuming that the lateral area is fixed. Indeed, this remnant magnetization changes with sample thickness despite being an intensive quantity, suggesting an extrinsic origin such as domain walls. We note that these models present possible means to interpret the changes to anomalous Hall conductivity between samples. However, we do not claim that these models and the resulting functional dependence of in-plane magnetization and out-of-plane Hall conductivity are the only possibilities. 

%Such samples are likely $AB$ or $AA$ samples with extrinsic origin of in-plane magnetization arising from domain effects as discussed above. 

\subsection{Density functional theory calculations} 

We performed DFT calculations  using the projector augmented-wave (PAW)~\cite{blochl_projector_1994,kresse_ultrasoft_1999} method as implemented in \textit{VASP}~\cite{Kresse_VASP}, with the potentials V\_sv, Nb\_sv and S provided by the VASP package.
In all calculations we employ the meta-GGA (R2SCAN)~\cite{furness2020accurate} as exchange-correlation functional. The cutoff energy for the plane-wave basis was set to 500 eV. We relax the structures of the AB polymorphs due to the large forces within the atoms with a force convergence criterion of $10^{-3}$ eV/\AA~and energy convergence criterion $10^{-6}$ eV. The density fo states (DOS) was calculated using the tetrahedron method with Blöchl corrections.

\begin{figure*}
    \centering
    \includegraphics[width=\linewidth]{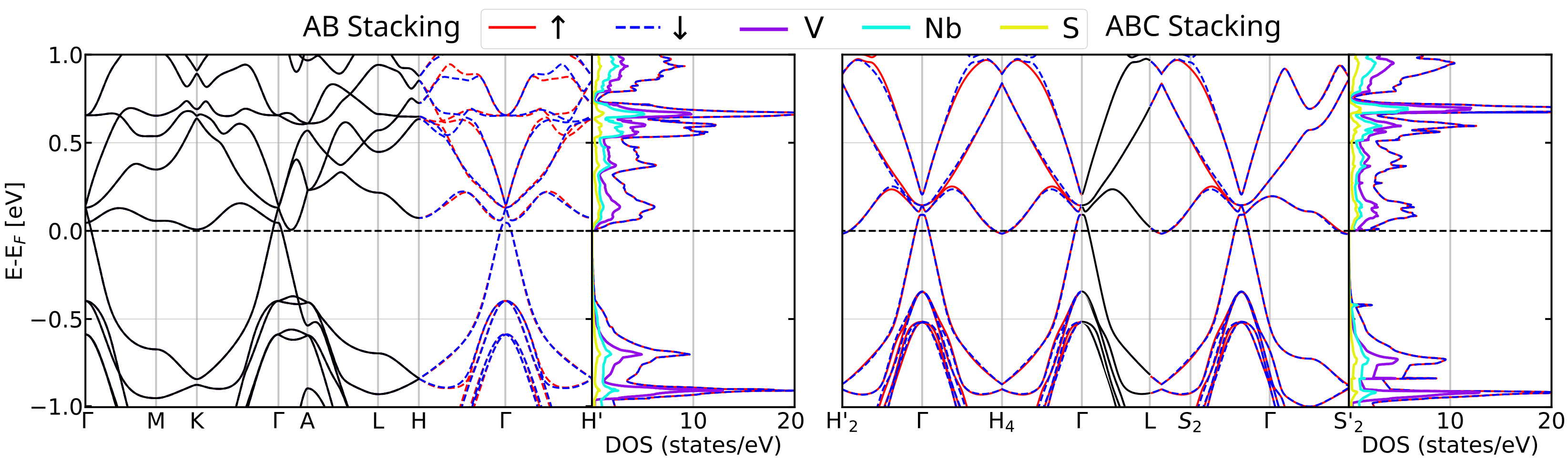}
    \caption{Spin resolved meta-GGA band structure and spin polarized and atomic species resolved density of states of \VNS for the AB and ABC polytypes. Along certain high symmetry lines the bands are degenerate (in black) and along other the bands are split (red and blue for spin up and spin down respectively). The inversion behavior typical of altermagnets can be clearly observed along the \textbf{H-$\Gamma$-H'} path in the AB stacking case. On the right of the band structure of each polytype, the DOS shows the AM behavior with the spin up and spin down populations being equal. In purple, cyan and yellow the atomic resolved DOS of \ce{V}, \ce{Nb} and \ce{S} respectively. The semimetallic behavior is shown with a low but non zero occupation around the Fermi energy.}
    \label{fig:DFT_bandstructure}
\end{figure*}

From our calculated spin-polarized electronic structure for both stacking structures, shown in Figure \ref{fig:DFT_bandstructure}, we observe that the DOS around and below the Fermi level is low and flat but non-zero, in agreement with the semimetallic experimental results of \VNS. Moreover, we obtain the overlap of the spin up and down DOS channels, resulting in a net zero magnetization, which is characteristic of the AM ground state of the material. We find the magnetic moment is \(1.9\mu_B\) per V, consistent with the local spin moment \(S=1\) expected for \(V\).

The AM behavior is visible in the spin-resolved  bands, showing spin splitting along certain high symmetry lines. 
For the AB stacking structure, that belongs to the \(P6_322\) magnetic group and is a 3\(^{\text{rd}}\)-phase non-centrosymmetric magnet, we observe the spin inversion between $\mathbf{H}$- $\bm{\Gamma}$ and $\bm{\Gamma}$ - $\mathbf{H'}$. 
This is consistent with the spin momentum locking being symmetric with respect to the inversion of \textbf{k}, despite the non-centrosymmetric crystal structure. 
For the ABC stacking structure belonging to the \(R\bar{3}c\) magnetic space group, the lower symmetry allows for spin splitting along more high symmetry lines. In this stacking there is also spin-splitting along the $\mathbf{H}$- $\bm{\Gamma}$- $\mathbf{H'}$ direction, as well as along $\mathbf{S_2}$- $\bm{\Gamma}$- $\mathbf{S_2'}$.

We compare the experimental parameters fit from experiment to the obtained from the DFT calculations. 
%Each vanadium atom is in the $V^{3+}$ oxidation state and is surrounded by six sulfur atoms, which form octahedra around it. Despite the slight angular distortions in the octahedra, the $V-S$ bond length remains constant. 
For that, we construct an effective Heisenberg Hamiltonian using the familiar form
\begin{equation}
    H = \sum_{ij} J_{ij} \vb{S}_i\cdot\vb{S}_j + D_z \sum_{i} {S_i^z}^2,
\end{equation}
where the magnetic exchange couplings $J_{ij}$ up to the 7$^{\rm th}$ are estimated using the total energy mapping analysis (TEMA)~\cite{Glasbrenner2015effect,kaib2022,riedl2022,razpopov2023}.
The total energies are estimated from spin polarized DFT calculations.
By using a least squares fit of the calculated energies to the spin-1 Heisenberg Hamiltonian we obtain the first seven nearest neighbor. For the AB stacking polytype we considered 25 different magnetic configurations in a cell $3\times2\times1$ times the conventional unit cell, whereas 16 were considered for the ABC polytype, in also a cell $3\times2\times1$ times the primitive unit cell.

%Due to the large forces in the AB polymorph crystal structure we relax this structure within DFT with force convergence criterion of $10^{-3}$ eV/\AA~and energy convergence criterion $10^{-6}$ eV.
%https://arxiv.org/pdf/2307.12366v3
The results of the calculated exchange parameters  for both polytypes are shown in Table \ref{tab:J_couplings_DFT}.
We find that both polytypes have FM intralayer interactions, which lead to a FM order within the layer, as seen in experiments. 
On the other hand, the interlayer interactions are in both polytypes dominated by AFM contributions, leading to an AFM interlayer order.
The single ion anisotropy term is found to be small in the order 0.01 meV, which is at the edge of DFT accuracy. This energy scale is agreement with the fit to the experimental neutron data as given in the main text.

\begin{figure}
    \centering
    \includegraphics[width=\linewidth]{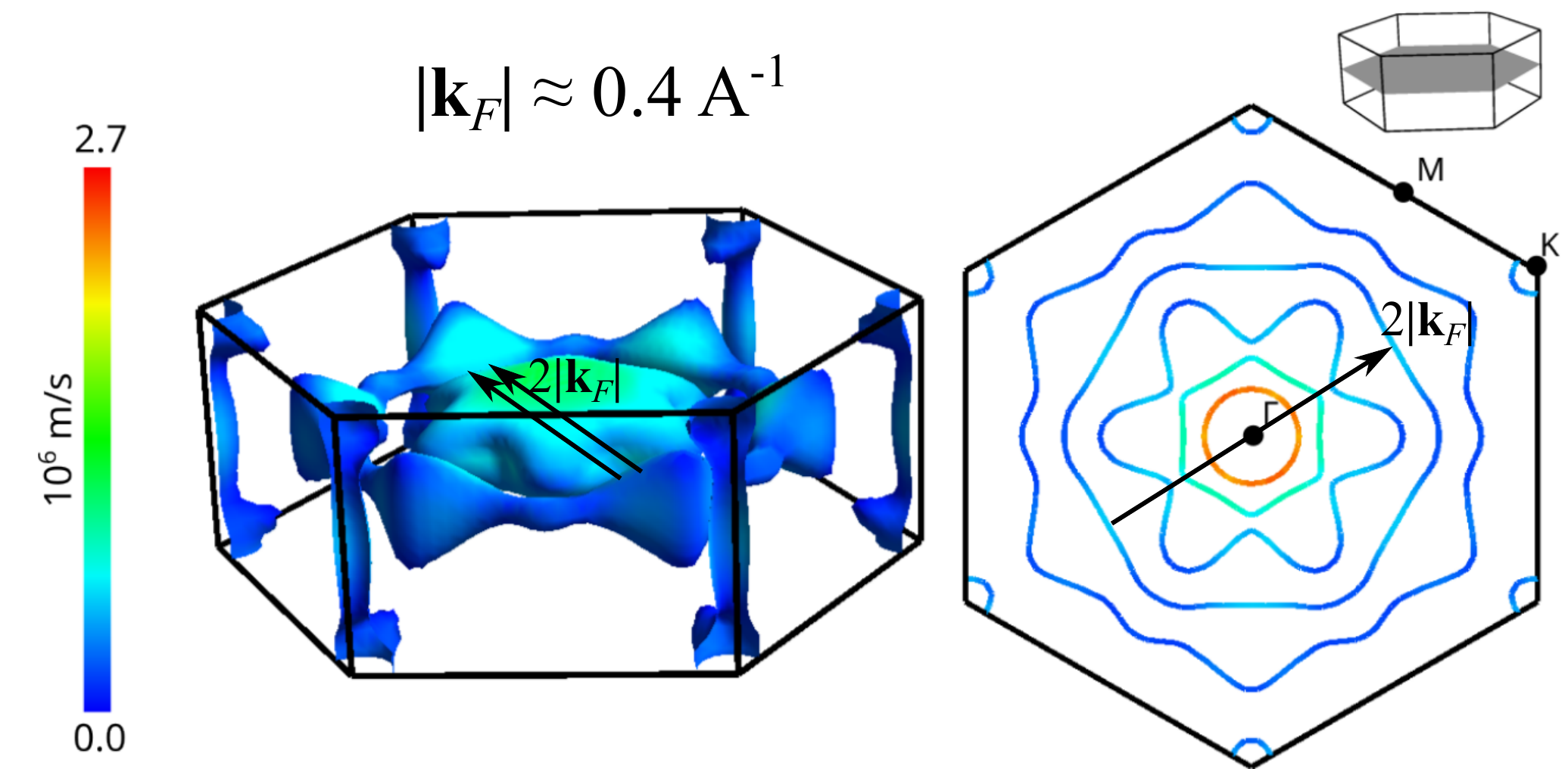}
    \caption{DFT meta-GGA calculations of the Fermi surface of \VNS with AB structure in the paramagnetic state showing the regions of the Fermi surface connected with a wavevector satisfying $|\vb{k}_F|\approx 0.4$~\AA$^{-1}$, compared with $k_F \approx 0.382(3)$~\AA$^{-1}$ from fits to the RKKY model. The color map indicates the Fermi velocity.}
    \label{fig:VNS_nesting}
\end{figure}

The Fermi surface Fig.\ref{fig:VNS_nesting} was obtained from paramagnetic DFT calculations using the full potential (linearized) augmented plane-wave and local orbitals method as implemented in WIEN2k~\cite{blaha2001wien2k}. All calculations within this basis were performed using the meta-GGA exchange-correlation functional. The charge and energy convergence criterion were set both to be \(10^{-5}\). The mesh was set to be \(19\times19\times8\) for the AB case and \(20\times20\times20\) for the ABC stacking case.

\end{document}